\documentclass[twocolumn]{aastex61}
\newcommand{\Bsixmaj}{0.037}
\newcommand{\Bsixmin}{0.022}
\newcommand{\Bsixpa}{67}
\newcommand{\Bthreemaj}{0.065}
\newcommand{\Bthreemin}{0.041}
\newcommand{\Bthreepa}{50.9}

\newcommand*\inst[1]{\unskip\hbox{\@textsuperscript{\normalfont$#1$}}}

\newcommand*\institute[1]{
  \begingroup
    \let\and\relax
    \renewcommand*\inst[1]{}%
    \renewcommand*\thanks[1]{}%
    \renewcommand*\email[1]{}%
  \endgroup
  \newcommand{\institutions}{#1}
}%

\let\oldarcsec\arcsec
\renewcommand\arcsec{\oldarcsec\xspace}%

\usepackage{latexsym}		% to get LASY symbols
\usepackage{graphicx}		% to insert PostScript figures
\usepackage{rotating}		% for sideways tables/figures
\usepackage{natbib}  % Requires natbib.sty, available from http://ads.harvard.edu/pubs/bibtex/astronat/
\usepackage{savesym}
\usepackage{amssymb}
\usepackage{amsmath}
\usepackage{morefloats}
\savesymbol{doublespace}
\usepackage{xspace}
\usepackage{color}
\usepackage{mdframed}
\usepackage{url}
\usepackage{subfigure}
\usepackage{grffile}
\usepackage{import}
\usepackage[utf8]{inputenc}
\usepackage{booktabs}
\usepackage{fancyhdr}

\usepackage[hang,flushmargin]{footmisc}
\usepackage{ifpdf}

\newcommand{\msun}{\ensuremath{M_{\odot}}\xspace}			%  Msun

\newcommand{\lsun}{\ensuremath{L_{\odot}}\xspace}			%  Lsun
\newcommand{\hh}{\ensuremath{\textrm{H}_{2}}\xspace}			%  H2

\newcommand{\water}{H$_{2}$O\xspace}		%  H2O
\newcommand{\kms}{\textrm{km~s}\ensuremath{^{-1}}\xspace}	%  km s-1

\def\ee#1{\ensuremath{\times10^{#1}}}
\newcommand{\degrees}{\ensuremath{^{\circ}}}
\newcommand{\perbeam}{\ensuremath{\textrm{beam}^{-1}}\xspace}

\def\eqref#1{Equation \ref{#1}}

\def\Figure#1#2#3#4#5{
\begin{figure*}[!htp]
\includegraphics[scale=#4,width=#5]{#1}
\caption{#2}
\label{#3}
\end{figure*}
}

\def\FigureOneCol#1#2#3#4#5{
\begin{figure}[!htp]
\includegraphics[scale=#4,width=#5]{#1}
\caption{#2}
\label{#3}
\end{figure}
}

\def\FigureTwo#1#2#3#4#5#6{
\begin{figure*}[!htp]
\subfigure[]{ \includegraphics[scale=#5,width=#6]{#1} }
\subfigure[]{ \includegraphics[scale=#5,width=#6]{#2} }
\caption{#3}
\label{#4}
\end{figure*}
}

\newenvironment{rotatepage}
{}{}

\def\FigureThree#1#2#3#4#5#6#7{
\begin{figure*}[!htp]
\subfigure[]{ \includegraphics[scale=#6,width=#7]{#1} }
\subfigure[]{ \includegraphics[scale=#6,width=#7]{#2} }
\subfigure[]{ \includegraphics[scale=#6,width=#7]{#3} }
\caption{#4}
\label{#5}
\end{figure*}
}

\def\FigureFour#1#2#3#4#5#6#7#8{
\begin{figure*}[!htp]
\subfigure[]{ \includegraphics[scale=#7,width=#8]{#1} }
\subfigure[]{ \includegraphics[scale=#7,width=#8]{#2} }
\subfigure[]{ \includegraphics[scale=#7,width=#8]{#3} }
\subfigure[]{ \includegraphics[scale=#7,width=#8]{#4} }
\caption{#5}
\label{#6}
\end{figure*}
}

\def\FigureFourVertical#1#2#3#4#5#6#7#8{
\begin{figure*}[!htp]
\subfigure[]{ \includegraphics[scale=#7,width=#8]{#1} }
\vspace{0.001mm} \\
\subfigure[]{ \includegraphics[scale=#7,width=#8]{#2} }
\vspace{0.001mm} \\
\subfigure[]{ \includegraphics[scale=#7,width=#8]{#3} }
\vspace{0.001mm} \\
\subfigure[]{ \includegraphics[scale=#7,width=#8]{#4} }
\vspace{0.001mm}
\caption{#5}
\label{#6}
\end{figure*}
}

\newcommand{\githash}{68a49ad}\newcommand{\gitdate}{2018-04-26\xspace}\newcommand{\sourcei}{SrcI\xspace}
\newcommand{\sourcen}{SrcN\xspace}
\newcommand{\sourcex}{SrcX\xspace}
\begin{document}
\newcommand{\nraojansky}{\affiliation{\it{Jansky fellow of the National Radio Astronomy Observatory, 1003 Lopezville Rd, Socorro, NM 87801 USA }}}
\newcommand{\nrao}{\affiliation{\it{National Radio Astronomy Observatory, 1003 Lopezville Rd, Socorro, NM 87801 USA }}}

\newcommand{\radboud}{\affiliation{\it{Department of Astrophysics/IMAPP, Radboud University Nijmegen, PO Box 9010, 6500 GL Nijmegen, the Netherlands}}}
\newcommand{\allegro}{\affiliation{\it{ALLEGRO/Leiden Observatory, Leiden University, PO Box 9513, 2300 RA Leiden, the Netherlands}}}
\newcommand{\casa}{\affiliation{\it{CASA, University of Colorado, 389-UCB, Boulder, CO 80309}} }

\newcommand{\berkeley}{\affiliation{\it{Radio Astronomy Laboratory, University of California, Berkeley, CA 94720}} }

\author[0000-0001-6431-9633]{Adam Ginsburg}
\nraojansky

\correspondingauthor{Adam Ginsburg}
\email{aginsbur@nrao.edu; adam.g.ginsburg@gmail.com}

\author{John Bally}
\casa

\author{Ciriaco Goddi}
\allegro
\radboud

\author{Richard Plambeck}
\berkeley

\author{Melvyn Wright}
\berkeley

\title{A Keplerian disk around Orion \sourcei, a $\sim15$ \msun YSO}
\begin{abstract}
   We report ALMA long-baseline observations of Orion Source I (\sourcei) with resolution
   0.03-0.06\arcsec (12-24 AU) at 1.3 and 3.2 mm.
   We detect both continuum and spectral line emission from \sourcei's disk.
   We also detect a central weakly resolved source that we interpret as a hot spot
   in the inner disk, which may indicate the presence of a binary system.
   The high angular resolution and sensitivity of these observations allow us
   to measure the outer envelope of the rotation curve of the \water
   $5_{5,0}-6_{4,3}$ line, which gives a mass $M_I\approx15\pm2$ \msun.
   We detected several other lines that more closely trace the disk, but were
   unable to identify their parent species.  Using centroid-of-channel methods
   on these other lines, we infer a similar mass.
   These measurements solidify \sourcei as a genuine high-mass protostar system
   and support the theory that \sourcei and the Becklin Neugebauer Object were
   ejected from the dynamical decay of a multiple star system $\sim$500 years
   ago, an event that also launched the explosive molecular outflow in Orion.
\end{abstract}

\section{Introduction}
Orion Source I (\sourcei) is the closest candidate forming high-mass ($M>8$ \msun) star, 
and as such is the most important protostar for testing basic theories
of how massive stars form.  However, despite its relative proximity at
a mere $\approx415$ pc from the sun \citep{Menten2007a,Kim2008a}, the mass of \sourcei
has been the subject of prolonged debate, with several estimates
putting
its mass below the classic 8 \msun threshold for a single star to go supernova
\citep[][]{Heger2003}.

Several attempts have been made to measure the mass of Orion \sourcei using the
rotation curve of various molecular lines:
\begin{itemize}
    \item \citet{Kim2008a} used 3D VLBI measurements of SiO masers  to infer a
        source mass $M=8$\msun.
        %by assuming the masers moving at orbital
        %velocity $v\sim12$ \kms came from a radius $r\sim42$ AU.
    \item \citet{Matthews2010a} used 3D VLBI measurements of SiO masers % velocities
        %at 0.5 mas (0.2 AU) resolution along a `bridge' hovering just above and
        %below \sourcei's disk, with an assumed radius $r\sim35$ AU and orbital
        %velocity $v=13.5$ \kms, 
        to infer a mass $M\approx8-10$ \msun.
    \item \citet{Hirota2014a} observed \water emission
        from the $v_2=0, 10_{2,9}-9_{3,6}$ and  $v_2=1, 5_{2,3}-6_{1,6}$ lines.
        %with $E_L=1846$ and 2939 K, respectively.
        %They imaged these lines
        %with $0.40\arcsec\times0.34\arcsec$ resolution and 
        They made
        velocity centroid maps of the position of peak intensity
        as a function of velocity to measure the rotationally supported
        mass in \sourcei.  They obtained a mass estimate of $5-7$ \msun
        %assuming a disk with radius 42-47 AU 
        using a model of a simple uniform
        ring orbiting at Keplerian velocity.
    \item \citet{Plambeck2016a} measured both the continuum SED and the rotation
        curve of gas around \sourcei.  They used a centroiding and modeling approach
        similar to \citet{Hirota2014a} to measure the rotation curve of SiO and
        vibrationally excited CO and infer the source mass $M\sim5-7$ \msun.
    \item \citet{Hirota2017b} used Si$^{18}$O J=12-11 to infer a mass $M=8.7$
        using a similar approach to \citet{Hirota2014a} and \citet{Plambeck2016a}.
        %\msun using an orbital velocity $v_{inner}=17.9$ \kms at  inner radius
        %$r_{inner} = 24$ AU.  These measurements are based on the same
        %centroid-of-velocity approach as \citet{Hirota2014a} and
        %\citet{Plambeck2016a}.
\end{itemize}

\sourcei's mass is an important parameter in models of the origin of the Orion
Outflow.  Several authors argue that \sourcei, BN, and \sourcen \citep[or,
alternatively, \sourcex;][]{Luhman2017a} were part of a single non-hierarchical
multiple system that underwent dynamical decay, and this decay somehow
triggered the outflow
\citep{Bally2005a,Rodriguez2005a,Goddi2011b,Moeckel2012b,Bally2011a,Bally2015a,Bally2017a,Rodriguez2017a}.
However, others have noted that the lower masses inferred for \sourcei above
are incompatible with this scenario
\citep{Chatterjee2012a,Plambeck2016a,Farias2017b}, which requires a mass $M_{I}
\gtrsim 15$ \msun.  An alternative scenario is described in which BN was
ejected from the Trapezium and had a close encounter with \sourcei that
triggered the outflow \citep{Tan2008a,Tan2008b,Chatterjee2012a}.  A third
alternative, that the outflow is driven by many independent sources
\citep{Beuther2008a}, is disfavored by the overall symmetry of the outflow
\citep{Bally2017a}.

We present new measurements of \sourcei's mass, finding it has $M_I \sim 15$ \msun.
In Section \ref{sec:observations}, we present details of the observations.
In Section \ref{sec:results}, we discuss measurements of the continuum
and spectral lines.  Section \ref{sec:discussion} discusses these results and
some of their simple physical implications.
We conclude in Section \ref{sec:conclusions}.
Several appendices present additional figures and detailed method discussion.

\section{Observations}
\label{sec:observations}

Observations were taken with two configurations in each of Band 3, 6, and 7 at
ALMA as part of project 2016.1.00165.S.  The epochs and broad details about the
configuration are given in Table \ref{tab:observations}.  The
multiconfiguration data were combined for all
images considered here.
The flux and phase calibrators are listed in Table \ref{tab:observations}.
Band 7 data are not discussed in this work because the data for
the long-baseline observations were not delivered by the time of submission; we
record the observational details here for completeness since they are part of
the same project in the ALMA archive.

Continuum images were produced with several weighting parameters to emphasize
different scales, though most of the discussion here will be limited to the
robust -2 weighted images with the highest resolution.  The calibrated data
delivered from the ALMA QA2 process were imaged directly, since we found
that self-calibration did not improve the image; we suspect the unmodeled, resolved-out
emission prevents us from obtaining good calibration solutions.

To emphasize the disk scales and eliminate ripple artifacts produced by
poorly-sampled large-scale structure in the map, we used data only from baselines
$>150$ m (115 k$\lambda$, angular scales $<1.8$\arcsec) in the robust -2 images
used for disk fitting and modeling.

\begin{table*}[htp]
\centering
\caption{Observation Summary}
\begin{tabular}{llllllll}
\label{tab:observations}
Date & Band & Array & Observation Duration &  Baseline Length Range  & \# of antennae & FluxCal & PhaseCal\\
     &      &       & seconds              & meters                    &              &         &         \\
\hline
08-Oct-2016 & 6 & 12m & 2332 & 17-3144 & 43 & J0522-3627 & J0541-0541 \\
31-Oct-2016 & 7 & 12m & 2671 & 19-1124 & 42 & J0522-3627 & J0532-0307 \\
19-Sep-2017 & 6 & 12m & 5556 & 41-12147 & 42 & J0522-3627 & J0541-0541 \\
24-Sep-2017 & 3 & 12m & 5146 & 21-12147 & 41 & J0423-0120 & J0541-0541 \\
25-Sep-2017 & 3 & 12m & 5180 & 41-14854 & 42 & J0423-0120 & J0541-0541 \\
\hline
\end{tabular}
\end{table*}

The continuum images have dynamic range in the vicinity of \sourcei of about 200-400.
These values are reported in Table \ref{tab:image_metadata} and are measured by
taking the ratio of the peak intensity in \sourcei to the standard deviation
within a neighboring, apparently signal-free region (an $r=0.7\arcsec$ circle 1.4\arcsec
to the northwest of \sourcei).

\begin{table*}[htp]
\centering
\caption{Continuum Image Parameters}
\begin{tabular}{ccccccccc}
\label{tab:image_metadata}
Band & Robust & Beam Major & Beam Minor & Beam PA               & $T_B$/$S_\nu$      & RMS & Source I $S_{\nu,max}$ & Dynamic Range\\
     &        & \arcsec    & \arcsec    & $\mathrm{{}^{\circ}}$ & $10^3$ K Jy$^{-1}$ & $\mathrm{mJy}~\mathrm{beam}^{-1}$ & $\mathrm{mJy}~\mathrm{beam}^{-1}$ & \\
\hline

B6 & -2 & 0.037 & 0.022 & 67.0 & 30.3 & 0.087 & 19.638 & 220 \\
B3 & -2 & 0.065 & 0.041 & 50.9 & 53.2 & 0.038 & 14.179 & 370 \\

\hline
\end{tabular}

\end{table*}

Spectral line image cubes were produced covering the complete data set to identify lines
associated with the disk.  Cubes were produced centered on \sourcei with robust
0.5 and -2 weightings.  These cubes were only cleaned in the
$0.5\arcsec\times0.5$\arcsec region immediately surrounding \sourcei, therefore lines with
significant emission from the surrounding medium may be significantly affected
by sidelobes.  Continuum-subtracted cubes were produced by subtracting
the median across the 1.8 GHz bandwidth in each spectral window. 
All cube analysis was performed using
\texttt{spectral-cube} (\url{https://spectral-cube.readthedocs.io/en/latest/}).

Relevant parameters of the cubes are described in Table \ref{tab:cube_metadata}.
For the noise estimate, we use the median absolute deviation (MAD) to estimate the
standard deviation over the full continuum-subtracted cutout cube, which
effectively ignores the few channels that have significant line emission (the
directly-measured standard deviation and MAD-estimated standard deviation
differ by $<5\%$).

\begin{table*}[htp]
\centering
\caption{Line Cube Parameters}
\begin{tabular}{cccccccccc}
\label{tab:cube_metadata}
Band & SPW & Freq. Range & Robust & Beam Major & Beam Minor & Beam PA               & RMS                               & RMS & Channel Width\\
     &     & GHz         &        & \arcsec    & \arcsec    & $\mathrm{{}^{\circ}}$ & $\mathrm{mJy}~\mathrm{beam}^{-1}$ & K   & \kms         \\
\hline

B3 & 0 & 85.463-87.337 & -2 & 0.066 & 0.043 & 45.0 & 2.5 & 144.8 & 3.4\\
B3 & 1 & 87.358-89.232 & -2 & 0.064 & 0.050 & 41.5 & 2.5 & 121.1 & 3.3\\
B3 & 2 & 97.462-99.336 & -2 & 0.061 & 0.039 & 45.8 & 2.0 & 108.4 & 3.0\\
B3 & 3 & 99.358-101.232 & -2 & 0.060 & 0.038 & 45.3 & 2.3 & 122.2 & 2.9\\
B6 & 0 & 229.168-231.042 & -2 & 0.026 & 0.022 & 64.0 & 2.8 & 115.9 & 1.3\\
B6 & 1 & 231.835-233.709 & -2 & 0.026 & 0.021 & 61.6 & 3.1 & 125.0 & 1.3\\
B6 & 2 & 214.277-216.151 & -2 & 0.027 & 0.023 & 62.8 & 3.2 & 131.5 & 1.4\\
B6 & 3 & 216.976-218.850 & -2 & 0.030 & 0.023 & 55.2 & 3.2 & 120.5 & 1.3\\
B3 & 0 & 85.463-87.337 & 0.5 & 0.101 & 0.072 & 40.6 & 0.8 & 17.3 & 3.4\\
B3 & 1 & 87.358-89.232 & 0.5 & 0.098 & 0.080 & 40.3 & 0.7 & 15.0 & 3.3\\
B3 & 2 & 97.462-99.336 & 0.5 & 0.091 & 0.060 & 43.6 & 0.7 & 16.8 & 3.0\\
B3 & 3 & 99.358-101.232 & 0.5 & 0.081 & 0.058 & 39.6 & 0.7 & 18.2 & 2.9\\
B6 & 0 & 229.168-231.042 & 0.5 & 0.043 & 0.035 & -88.1 & 0.9 & 14.6 & 1.3\\
B6 & 1 & 231.835-233.709 & 0.5 & 0.043 & 0.034 & -87.4 & 1.0 & 15.7 & 1.3\\
B6 & 2 & 214.277-216.151 & 0.5 & 0.046 & 0.037 & -88.7 & 1.2 & 18.2 & 1.4\\
B6 & 3 & 216.976-218.850 & 0.5 & 0.049 & 0.039 & 72.7 & 1.0 & 14.1 & 1.3\\

\hline
\end{tabular}

\end{table*}

Cutouts of the data used for the analysis in this work along with the software and scripts
used for the analysis are presented at \url{https://zenodo.org/record/1213350}.

\section{Results}
\label{sec:results}

\subsection{Continuum}
We detect the disk in the continuum at 3.2 mm, 1.3 mm, and 0.8 mm (Figure
\ref{fig:continuum_data_B6} shows the 3.2 and 1.3 mm images\footnote{At the
time of submission, the long-baseline 0.8 mm data products had not been
delivered, so they are excluded from the analysis presented here.}).  At 1.3
mm, where we have enough resolution to clearly distinguish the line-emitting
region from the disk midplane, we detect spectral lines only from the surfaces
above and below the continuum disk (Figure \ref{fig:U1peak}).  The nondetection
of lines in the disk midplane is a strong indication that the continuum is
optically thick, as has previously been noted \citep[e.g.,][]{Plambeck2016a}.

We fit the highest-resolution 1.3 mm and 3.2 mm continuum image with a simple
model to determine the basic
observational structure.
The optimization was performed using a Levenberg-Marquardt
fitter \citep{Newville2014a}. 
We used a linear model (i.e., an infinitely thin
perfectly edge-on disk) for the disk, with endpoints and amplitude as free parameters.
This simple model left significant residuals, so we added a two-dimensional
Gaussian smoothing kernel as another three free parameters to obtain a
substantially better fit.
The models and their residuals
are shown in Appendix \ref{appendix:contmodel}.

We determined that the disk is resolved in both
directions, with a vertical FWHM height of about 20 AU and a length of about
100 AU (Table \ref{tab:continuum_fit_parameters}).  These measurements
are close to those published by \citet{Plambeck2016a}, though their data
only marginally resolved the source at wavelengths 1.3 mm and shorter.

This simple model leaves a significant residual compact source near the center
of the disk, which we measured by adding a smeared point source to the model
(see Appendix \ref{appendix:contmodel}).  We have allowed the source to be
smeared only in the direction of the disk's elongation, requiring only two
additional free parameters.  This source is discussed further in Section
\ref{sec:ptsrc}.

Table \ref{tab:continuum_fit_parameters} lists the fitted parameters.  It 
includes measurements of the total integrated intensity recovered in the model
and the ratio of the compact central source to the total.  We also display
fits to the \citet{Reid2007a} 7 mm continuum data with the same model;
these fits do not contain any absolute astrometry information.

\begin{table*}[htp]
\centering
\caption{Continuum Fit Parameters}
\begin{tabular}{ccccccccccc}
\label{tab:continuum_fit_parameters}
Frequency & Disk FWHM & Disk Radius & Disk PA & Pt RA & Pt Dec & Pt Amp & Pt Width & Pt Flux & Total Flux & Pt \% \\
$\mathrm{GHz}$ & $\mathrm{AU}$ & $\mathrm{AU}$ & $\mathrm{{}^{\circ}}$ & $\mathrm{s}$ & $\mathrm{{}^{\prime\prime}}$ & $\mathrm{mJy}$ & $\mathrm{AU}$ & $\mathrm{mJy}$ & $\mathrm{mJy}$ & $\mathrm{}$ \\
\hline
43.165 & 25 $\pm$ 0.44 & 41 $\pm$ 1.1 & -36 & - & - & 1.00 $\pm$ 0.01 & 16 $\pm$ 0.69 & 3 & 10 & 29\% \\
93.3 & 17 $\pm$ 0.09 & 37 $\pm$ 0.71 & -38 & 0.518 & -0.0405 & 2.00 $\pm$ 0.03 & 9.4 $\pm$ 0.38 & 5.7 & 57 & 10\% \\
224.0 & 21 & 51 $\pm$ 0.78 & -37 & 0.518 & -0.0409 & 3.20 & 23 & 15 & 280 & 5.5\% \\
\hline
\end{tabular}

\par The pointlike source position is given as RA seconds and Dec arcseconds offset from ICRS 5h35m14s -5d22m30s.  The error on this position is 0.003s (RA) and 0.0003\arcsec (Dec). For the 7 mm data, the position is left blank because we do not have astrometric information for those data (they were self-calibrated on a bright maser whose position was not well-constrained).  The disk FWHM is the vertical full-width half-maximum of the fitted Gaussian profile.  No formal parameter errors were measured for several of the 224.0 GHz fitted quantities because of a linear algebra failure in the fitter; the errors are likely similar to the 93.3 GHz fit errors. The Pt Flux and Total Flux columns report the integrals of the best-fit models.
\end{table*}

The disk position angle points to within 2 degrees of the Becklin-Neugebauer
object (Orion BN); the PA of the vector from \sourcei to Source BN is -37.6
degrees, while the measured disk position angle is -36 to -37 degrees.
This coincidence was noted by \citet{Bally2011a} and \citet{Goddi2011b}.

The disk has a peak brightness temperature at 1.3 mm of $\sim600$ K at the
position of the compact source and $\sim400-500$ K at other positions, with a
gradual decline from the center to the exterior.  These measurements agree with
the continuum model of \citet{Plambeck2016a}, who inferred the presence of an
optically thick $T=500$ K surface from the SED.  The 3.2 mm continuum has a
higher peak brightness temperature at the position of the central compact
source, but otherwise is consistent with the 1.3 mm brightness (see Figure
\ref{fig:continuum_data_B6}).

\FigureTwo
{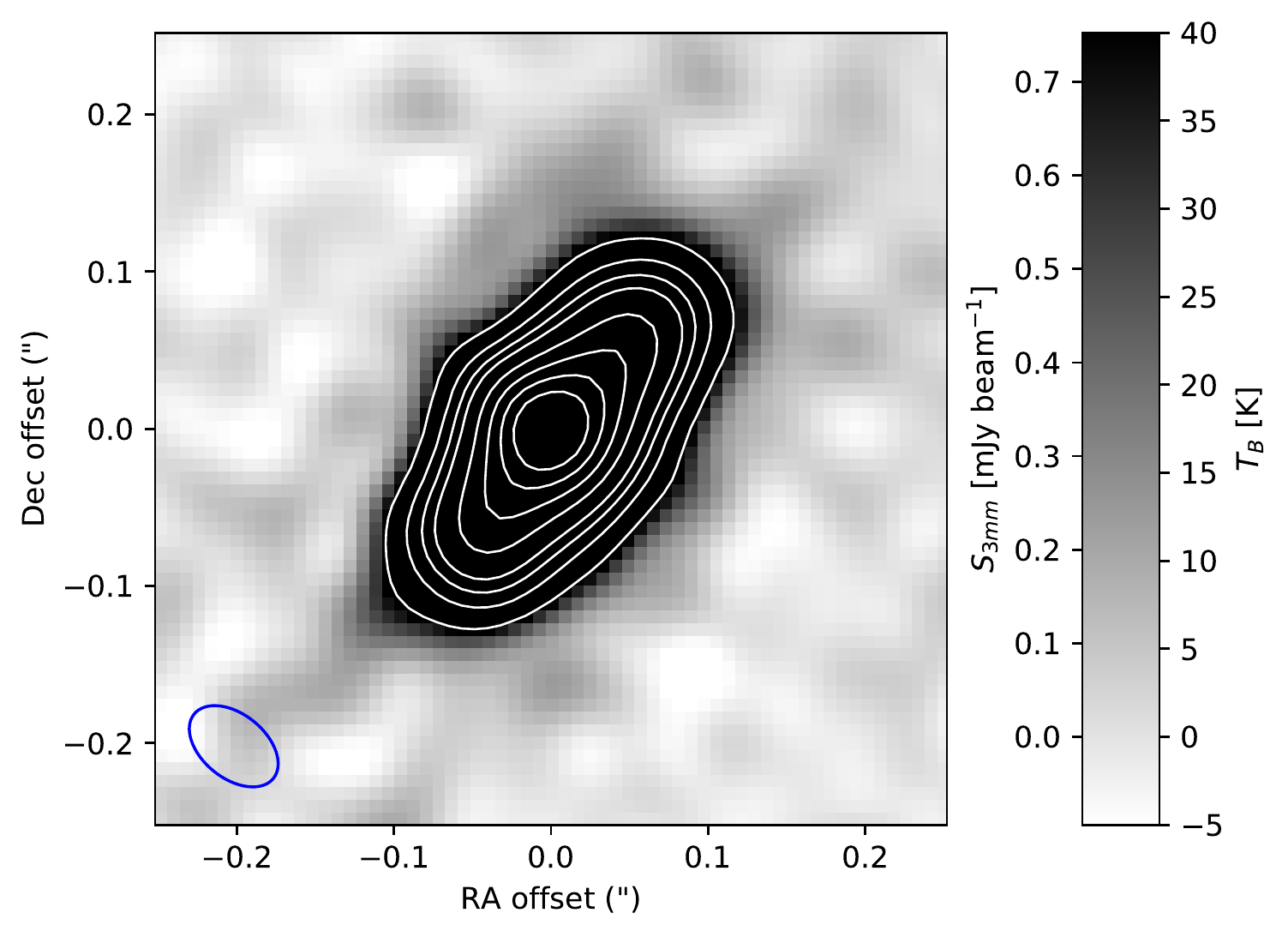}
{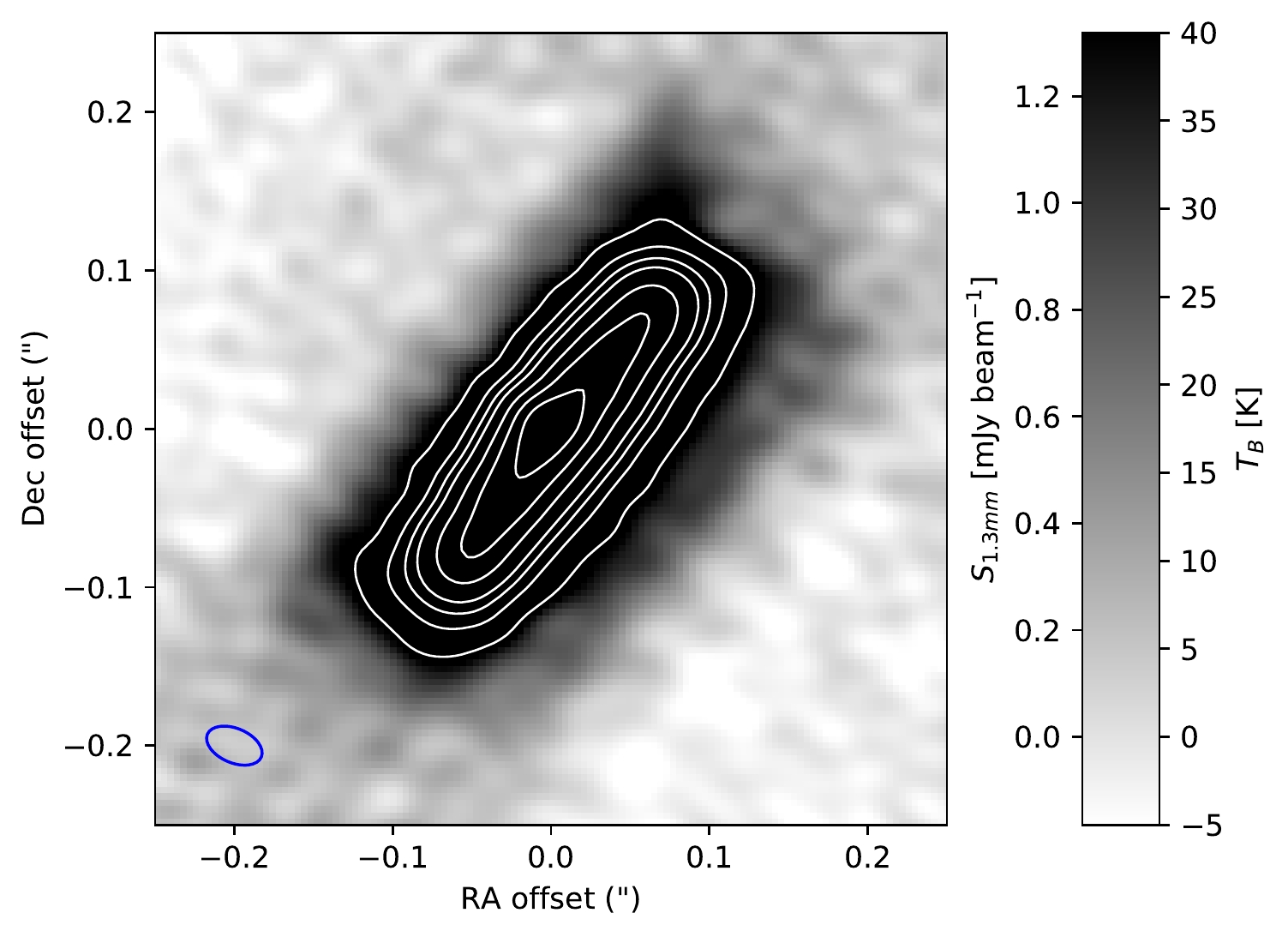}
{The robust -2 continuum image of Orion \sourcei at 3.2 mm (left) and 1.3 mm (right).
The beam is shown
in the bottom left, with size $\Bthreemaj\arcsec\times \Bthreemin\arcsec$ at
PA$=\Bthreepa\degrees$ (3.2 mm, left) and $\Bsixmaj\arcsec\times \Bsixmin\arcsec$ at
PA$=\Bsixpa\degrees$ (1.3 mm, right).
Contours are overlaid at $T_B=$50, 100, 150, 200, 300, 400, and 500 K.
The displayed coordinates are offsets from ICRS 05:35:14.5172 -05:22:30.612
(3.2 mm) and ICRS 05:35:14.5173 -05:22:30.6135 (1.3 mm).
}
{fig:continuum_data_B6}{1}{3.5in}

\subsection{SiO Lines}
We detect several SiO lines, including the 1 mm $^{28}$SiO v=0 and v=1 J=5-4 lines,
the the 3 mm isotopologue lines $^{29}$SiO v=0 and v=1 J=2-1, and the 3 mm
$^{28}$SiO v=0 and v=1 J=2-1 lines.  Several of these transitions are known and
well-studied masers.  Some images of these data are shown in
Appendix \ref{sec:siopv}, but because the emphasis of this work is not on the
outflow, we do not discuss the SiO further here.

\subsection{Water Line}
The next brightest line after the masing SiO lines is the \water
$5_{5,0}-6_{4,3}$ line at 232.68670 GHz, with $E_U=3461.9$ K.
\citet{Hirota2012a} detected this line in 2\arcsec resolution ALMA Science
Verification data, but believed it to be masing.  We report here that, because
it is similar in morphology and excitation level to the 336 GHz vibrationally
excited water line reported in \citet{Hirota2014a}, and it has a peak
brightness temperature $\sim1500\pm100$ K (in the robust -2 maps; see Figure
\ref{fig:U1peak}), it is most likely a thermal line.

The water line traces an X-shaped feature above and below the disk, resembling
the overall distribution of SiO masers.  The water is not directly aligned with
the continuum disk (Figures \ref{fig:U1peak} and \ref{fig:h2omom0}), but it
does exhibit emission parallel to the disk at small ($<20$ AU) separation.  
The morphology of this line confirms that it traces both the disk and
the inner rotating outflow discussed by \citet{Hirota2017b} \citep[see
also][]{Kim2008a,Matthews2010a}.

Because the water emission is thermal, it exhibits less extreme
brightness fluctuations across the image than the SiO masers, allowing us to
fit an upper-envelope velocity curve in Section \ref{sec:kinematics}.
The kinematics of the water line away from the disk are discussed
further in Appendix \ref{appendix:waterlinerevisited}.

\subsection{Other lines}
\label{sec:otherlines}
Several unidentified lines are observed in  emission at the outer edge of the
continuum disk.  A table of their approximate rest frequencies is presented
in Appendix \ref{appendix:stackedspectra}.  They all share a common morphology,
though they vary in strength.  The peak signal from these lines appears around
the $T_B\sim150$ K contour in the robust -2 weighted 1.3 mm continuum images
(Figure \ref{fig:U1peak}), and the lines are particularly strong at the
endpoints of the disk.  Little line emission is detected where the continuum is
brightest, $T_B\gtrsim300$ K.

The best explanation for these lines is that they trace the outer surface of a
mostly optically thick (in the continuum) disk.  In this scenario, the lines
have an excitation temperature similar to the brightness temperature of the
disk, but have optical depths of order $\tau\sim0.1-1$.  Directly toward the
disk continuum emission peak, since the line excitation temperature is the same
as the background continuum temperature, $T_{ex}=T_{bg}$, no emission (or
absorption) is observed.  Just above and below the disk midplane continuum
peak, the dust column density (and therefore optical depth) drops rapidly, but
the molecular optical depth drops more slowly, so some emission is still
observed ($T_{ex}\sim200-500$ K, but $\tau_{line}\sim0.1-0.5$, resulting in the
$T_{B,max} \sim 100$ K observed).  At the disk endpoints, the column density of
molecular material is higher because we are looking along the tangent of the
disk, so the line optical depth and therefore brightness are greater.

Since these lines only appear immediately around the disk, and in particular
because they peak just outside of the dust emission along the disk axis, they
are the most direct tracers of the disk's kinematics.

\FigureThree
{{f3}.pdf}
{{f4}.pdf}
{{f5}.pdf}
{Peak intensity map of an unknown line (U230.322; left) and the \water
line (middle, right) with continuum overlaid
in red contours at levels of 
50,  300, and 500 K. %robust -2
White contours are shown at 50, 100, and 150 K (left), and  500, 750, 1000, and 1250 K
(middle, right).
The \water and unknown line clearly trace different physical structures, as
they exhibit no coincident emission peaks.  The water line does
not exhibit emission directly along the disk midplane.
The left two figures are robust 0.5 weighted images, while the right
is a robust -2 weighted image with higher resolution and poorer
sensitivity.  The U230.322 line is not detected in the robust -2 cubes.
The positions shown are offsets from coordinate ICRS 05:35:14.5184 -05:22:30.6194.
}
{fig:U1peak}{1}{2.4in}

\subsection{Kinematics: a Keplerian disk}
\label{sec:kinematics}
The disk appears to exhibit a Keplerian rotation curve, which allows us to use
the velocity profile to measure the central mass.
Following \citet{Seifried2016a}, we measure the outer edge of the detected
emission in a position-velocity (PV) diagram to define the rotation curve
surrounding \sourcei. 
The emission along each line-of-sight
in the PV diagram is followed from
its peak down to some emission threshold.
We use a threshold of 5-$\sigma$, as was done in \citet{Seifried2016a}, but we
also assess the importance of this threshold in Section
\ref{sec:errorestimate}.
Additionally, to facilitate direct comparison with
previous works, we use the centroid-of-velocity-channel approach
in Appendix \ref{appendix:centroids} (although 
\citet{Seifried2016a} warn it may underestimate the central mass)
and obtain similar results, albeit from different spectral lines.

Of the detected lines, the \water line spans the widest range of
velocities as a function of radius.  As shown in Figure \ref{fig:h2okepler},
there is \water emission spanning at least radii $r\approx 10$ to 100 AU.  Many
other molecules, most of
which we have not been able to identify, span a range of radii 30-80 AU, while SiS
spans 30 AU to an unconstrained outer radius.   The SiO lines span varying radii,
but they are predominantly detected far from the disk midplane (however,
see Appendix \ref{sec:siopv}, in which SiO also exhibits Keplerian velocity
curves).

We find that a 15 \msun edge-on Keplerian rotation curve fits\footnote{We do
not report best-fit parameters and statistical errors here because the errors
on the outer envelope are poorly characterized and likely dominated by
systematic errors such as channel discretization.} the outer edge of
the \water line well (Figure \ref{fig:h2okepler}).  The 15 \msun profile is
also an acceptable match to the outer profiles of the unidentified lines
described in Section \ref{sec:otherlines} and shown in supplemental figures in
Appendix \ref{sec:ulinefigures}.

A lower-mass central source is consistent with the \water data only if we allow
for substantial line broadening driven by turbulence.  We estimate an upper
limit on the turbulent line broadening of $FWHM\lesssim4$ \kms based on the
narrowest features observed in the \water and other unknown lines, which
results in a one-sided broadening of $HWHM<2$ \kms; if such line-broadening is
present, the mass may be lower by $\sim20-30\%$.
However, since \citet{Flaherty2017a} observe stringent upper limits on
turbulence in lower-mass disks, we expect turbulent broadening to be relatively
small, and possibly negligible.  The upper-limit line broadening we observe is
both consistent with line blending from unresolved kinematics within the beam
and is close to the intrinsic velocity resolution of our data.

A substantially smaller mass, such as the 5-10 \msun suggested previously
\citep{Plambeck2016a,Hirota2014a}, is inconsistent with the data: for models
with such masses, emission is clearly detected outside of the predicted
Keplerian curve (see, e.g., the green curve in Figure \ref{fig:h2okepler}).  

\Figure
{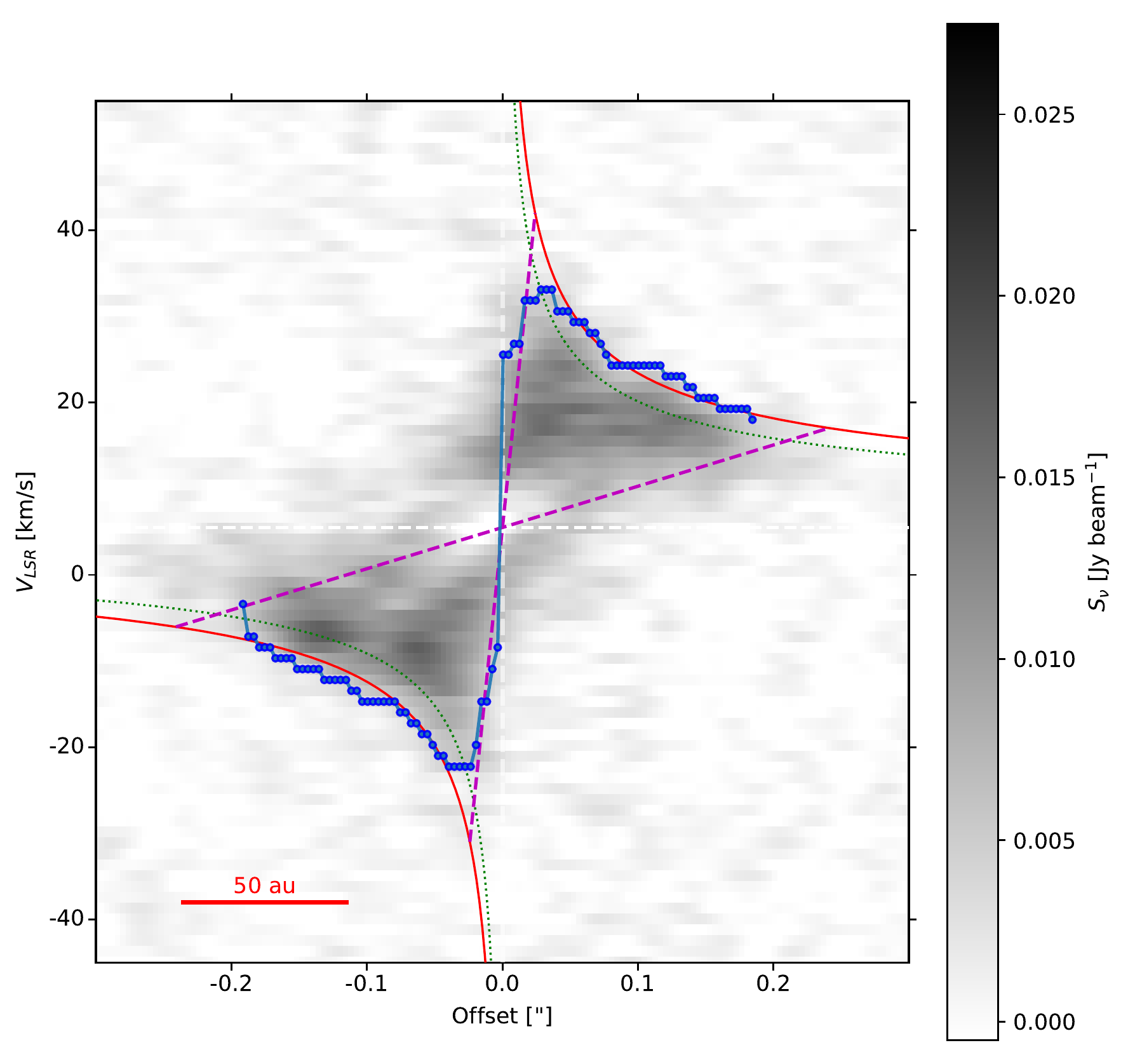}
{Position-velocity diagram of \water $5_{5,0}-6_{4,3}$.
The colorbars show average intensity along the extracted region in units
of mJy \perbeam.
The blue line with dots is the outer envelope of the velocity curve
determined using the method of \citet{Seifried2016a}.
The red solid and green dotted curves show the Keplerian velocity profile
surrounding a 15 and 10 \msun central source, respectively.
White dashed lines indicate the adopted source central position
05h35m14.5172s -05d22m30.618s (ICRS)
and central velocity (5.5 \kms).
The purple dashed lines show the full orbital path for radii of
10 and 100 AU and indicate the approximate limits of the disk.
This PV diagram is extracted from the midplane of the robust 0.5 image,
but the emission displayed is beam-smeared from just above and below the
continuum disk.
}
{fig:h2okepler}{1}{7in}

\subsubsection{Examination of alternative velocity profile models}
To support the argument that the velocity profiles are Keplerian, as opposed to
some other power-law profile as has been found for the outer envelopes of
several low-mass YSOs
\citep{Lee2017h,Aso2017b,Ohashi2014a,Lindberg2014a,Murillo2013a}, we show
power-law fits to the outer envelope velocity profile of the \water line
in Figure \ref{fig:h2opowerlaws}.
This figure convincingly demonstrates that
a power law $\alpha=1$ \citep[e.g., as observed in the outer parts of low-mass
YSO disks;][]{Aso2017b} is inconsistent with the data.  While the best-fit
profile of $\alpha\approx0.4$ is slightly shallower than the Keplerian
$\alpha=0.5$ curve, both shallow profiles are consistent with the data.
The limited spectral resolution of our data is evident in this plot, where
there is little separation in velocity from $\sim30-50$ AU; higher
spectral resolution observations or more sophisticated modeling may be able to
provide a tighter constraint on the power-law slope.

\FigureOneCol{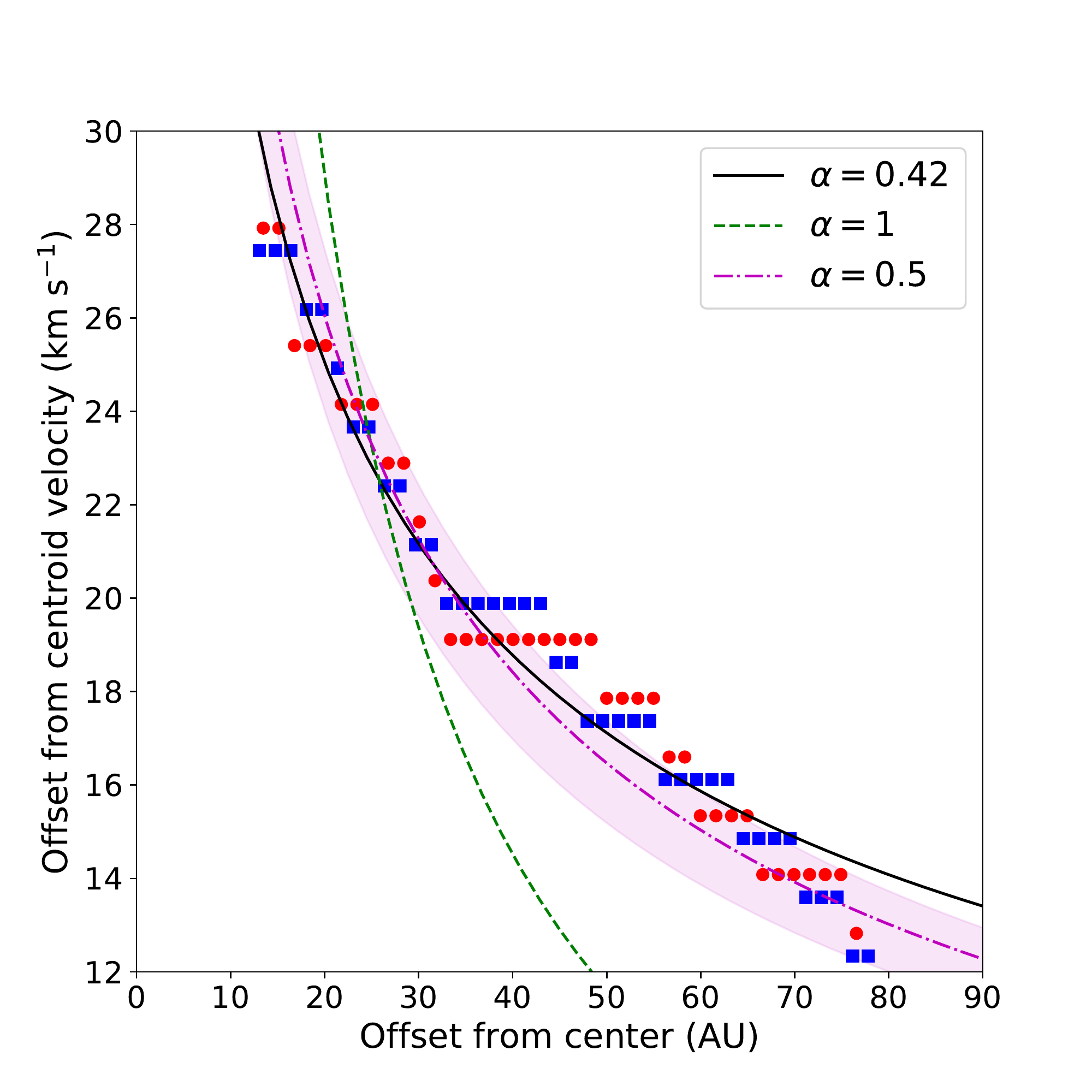}
{Radial profile of the outer-envelope velocity profile extracted from
Figure \ref{fig:h2okepler}b.  The red and blue points represent the redshifted
and blueshifted components of the velocity profile, respectively.  The velocities
are shown relative to the best-fit centroid velocity for the \water line,
$v_{LSR}=5.2$ \kms.  The curves show the best-fit power-law (black solid line)
and the best-fit curves with fixed powerlaw indices of 0.5 (magenta dotted line)
and 1 (green dashed line).  The magenta filled curve shows a powerlaw index
$\alpha=0.5$, i.e., a Keplerian rotation curve, for the range $13 \msun < M <
17 \msun$. }
{fig:h2opowerlaws}{1}{3.0in}

The mass for the $\alpha=0.5$ curve is $M=15$ \msun.  We do not determine
masses for the other models since they are not consistent with a pointlike
gravitational potential.  We show the curves for a central 13-17 \msun source
in filled magenta: since there are many points above the curve, a more massive
central source is plausible, while a less massive source is unlikely.

\subsubsection{An estimate of the error on the mass measurement}
\label{sec:errorestimate}
We  assess the uncertainty introduced by the threshold level adopted in the
velocity envelope profile measurement.  Figure \ref{fig:seifreidthresholderror}
shows the effects of increasing or decreasing the threshold, which is to
decrease or increase the measured mass, respectively.  These figures suggest
that our measurement uncertainty with the PV envelope fitting technique  is
approximately 2 \msun.

\Figure{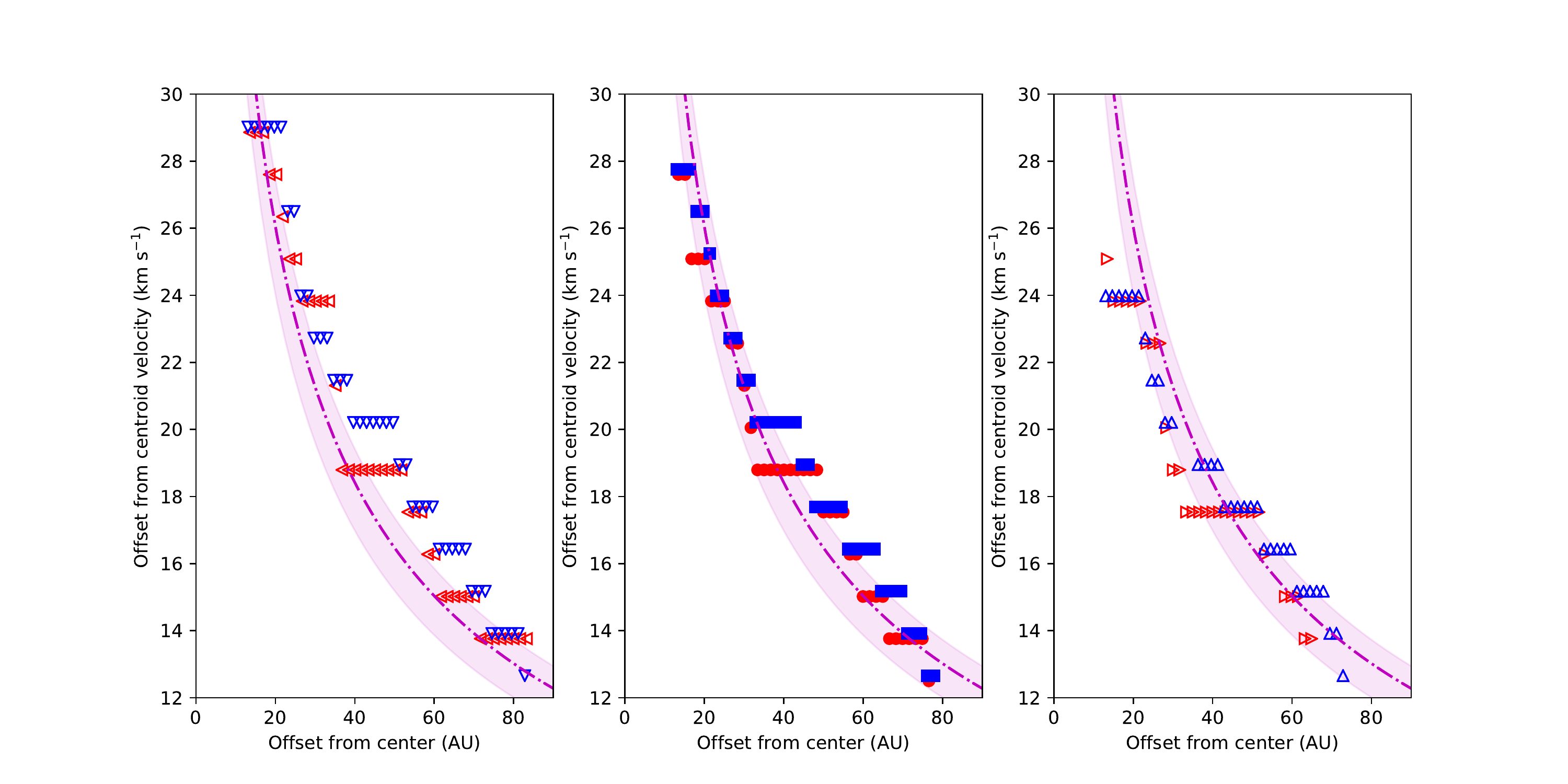}
{Demonstration of the effect of a changing threshold in the Seifried method.
The panels show a lower 3-$\sigma$ threshold (left), the adopted 5-$\sigma$
threshold (center), and a higher 7-$\sigma$ threshold (right).  
The magenta highlighted region is the same 13-17 \msun Keplerian curve shown
in Figure \ref{fig:h2opowerlaws}.  
}
{fig:seifreidthresholderror}{1}{7.5in}

\section{Discussion}
\label{sec:discussion}
\subsection{The mass of \sourcei}
We measure a mass for the object at the center of the disk of $M_I=15$
\msun, which is higher than most measurements previously reported.  Our mass
measurement is higher than previous works in part because our spatial
resolution is high enough to allow a direct fit of the rotation curve to the
outer envelope of an emission line in position-velocity space.
Additionally, though, the greater sensitivity of these observations allowed
us to detect the outer envelope of the \water position-velocity diagram
and detect - and resolve - several unknown lines that directly trace the disk.
The inconsistency between these new estimates and the lower masses previously
derived from SiO measurements hints that, in this system, SiO chemically
selects a kinemetically distinct region from the disk.

Even if \sourcei consists of an equal-mass binary, this mass measurement
confirms that the Orion Molecular Cloud is presently a region with ongoing
high-mass star formation.

\subsection{The luminosity of \sourcei}
Since we observe an optically thick surface, we can infer the luminosity
required to keep such a surface at the observed $T_{B,1.3 mm}\approx$500 K assuming
it is heated only by radiation.  Taking the disk radius to be 50 AU,
the required central source luminosity is 6500 \lsun.  This estimate
should be taken as a lower limit, since the inner disk is likely to be
optically thick and capable of shielding the outer disk, thereby
keeping the observed $\tau=1$ surface at 1.3 mm cooler than would
be produced by radiative equilibrium with the central star.

\subsection{Properties of the disk}
Our observations yield disk properties nearly identical to those in
\citet{Plambeck2016a}, so we do not revisit their disk mass or density
estimates.  We note, however, that these new observations have sufficient
angular resolution to distinguish the molecular lines that trace the outflow
from those that directly trace the disk.

\subsection{The dynamical decay scenario}
Several authors \citep[][]{Gomez2008a,Goddi2011b,Bally2011a} suggested that
the high proper
motion of \sourcei, BN, and \sourcen, combined with the observed \hh outflow,
implied the outflow and the runaway stars were produced in the same single
event $\sim500$ years ago.  That event was the dynamical decay of a
non-hierarchical multiple system, i.e., it was the interaction of multiple
stars at the center of a small cluster.  More recent observations by
\citet{Luhman2017a} have shown that \sourcen is unlikely to have participated
in this interaction, but instead that \sourcex, another star within the same
field, has high proper motion that points back to the interaction center (Bally et al,
in prep).

\citet{Farias2017b} report that, while any dynamical decay scenario involving
Sources I, X, and BN that can reproduce the observed proper motions are
unlikely, those with a higher mass for \sourcei ($M_I>14$ \msun) are the only
ones capable of producing the observed proper motions\footnote{Their results
are similar to those obtained in \citet{Goddi2011b} and \citet{Moeckel2012b},
but now with \sourcex instead of \sourcen as the third member of the
interaction.}.  Our observed higher mass for \sourcei, $M_I\gtrsim15$ \msun,
therefore implies that the
dynamical interaction scenario remains viable.

\subsection{Is the disk consistent with the dynamical ejection model?}
\citet{Plambeck2016a} argue that both the mass of \sourcei and the presence of
the disk rule out the dynamical ejection model of \citet{Bally2011a}.  We have
shown that the star is significantly more massive, but what about the disk?

Following \citet{Bally2011a}, we note that the disk truncates at $R<50$
AU.  At this radius, the orbital timescale is $\sim70$ years, so gas at the
disk's outer radius would have had five to ten dynamical times to relax into a
circular disk configuration after the explosive event.

The alignment of \sourcei's disk with the I-BN vector is consistent with a
dynamical interaction between these sources.  If the ejection resulted in
\sourcei and BN being launched in nearly opposite directions from their center
of mass (which must have been moving in the rest frame of the Orion nebula; 
Bally et al. in prep), any material around \sourcei that remained bound would
be dragged in the direction of \sourcei, and would therefore have a resulting
angular momentum vector orthogonal to the direction of motion.  Any material
with
velocity relative to \sourcei 
$$v < v_{esc} = 23 \kms (M_I/15\msun)^{1/2}  (r/50~\mathrm{AU})^{-1/2}$$
would remain bound.
Assuming \sourcei's present-day proper motion of 11.5 \kms reflects its velocity
at the time of ejection, less than half of the original disk mass would
have been lost, while the rest would remain bound (material moving in the
direction directly opposite \sourcei's ejection direction would have net
velocity relative to \sourcei high enough to escape; the greatest mass
loss would occur if the disk was already in the direction of \sourcei's eventual
launch).
Material outside $R\gtrsim200$ AU (where $v_I=11.5$ \kms $> v_{esc}$) would
likely all have become unbound, while other material would be retained in a
disk parallel to Source I's proper motion.

\subsection{The compact source in the disk}
\label{sec:ptsrc}
We have detected a compact (but marginally resolved) source near the center of
the disk at both 3.2 mm and 1.3 mm.  The source has a spectral energy distribution
that is shallow from 3.2 mm to 1.3 mm ($\alpha\approx1.1$).  Since the source is nearly
coincident with an edge-on disk that we show is optically thick at 1.3 mm, it is
likely that the source is thermal but is significantly attenuated by the disk
at 1.3 mm and seen with less attenuation at 3.2 mm.

\citet{Reid2007a} used comparable-resolution 7 mm VLA data to infer
the presence of a 2.2 mJy source at the center of the \sourcei disk.
The compact-source-to-disk flux ratio at 7 mm was $\sim30\%$, substantially
higher than we observe at 1.3 mm and somewhat
higher than at 3.2 mm (Table \ref{tab:continuum_fit_parameters}).  
The spectral index of this compact source from 7 to 3.2 mm is $\alpha=1.6$,
approaching that of an optically
thick blackbody.

The central source has a surface temperature $T\geq1250$ K, the brightness
temperature of a 2.2 mJy source within a $41\times28$ milliarcsecond beam at 7
mm.  If the source is a 5000 K spherical blackbody \citep[e.g.,][]{Testi2010a},
it must have a radius $R=7.5$ AU.  Such a gigantic star is implausible, as it
would produce a luminosity of 1.5\ee{6} \lsun, several orders of magnitude
higher than the total luminosity in the region.  We therefore argue
that this emission source is not a star.

What is the emission mechanism from this central source?
It could simply be hot, optically thick dust that is partly obscured by the
cooler disk at higher frequencies.  The  extension of this `source' along the
disk direction (Appendix \ref{appendix:contmodel}, Figure
\ref{fig:contmodel_residuals_B6}) suggests that we are
seeing the hot inner disk.  As pointed out by \citet{Plambeck2016a}, it is
quite unlikely to be classical free-free emission from protons
and electrons, since there are no detected
recombination lines.  However, it is still plausible that the emission is
produced by brehmsstrahlung emission from HI and \hh
\citep{Reid2007a,Baez-Rubio2018a}.

The source is slightly offset from the center of the disk by $5.8\pm1.5$ AU in
projection\footnote{We measure the errors on the source position by fitting a
2D Gaussian model to an image with the disk model subtracted.  While this approach
yields a useful statistical error, it does not account for the systematic error
introduced by fitting the disk model. }.  This offset,
combined with the source's extent, implies that it is
not a single central source, but
instead is a hot region of the inner disk.  Such an asymmetry in the disk could
be driven either by instability in the disk or, if the central star is a
binary, by the proximity of the more luminous companion.

If this source is an inner edge of the disk, it may imply the presence of a
binary that has cleared the area within $r<6-10$ AU.  Since a tight binary is
one of the expected outcomes of the dynamical interaction scenario
\citep{Goddi2011b}, this
detection of the inner region in dust emission provides additional
circumstantial evidence for that scenario.

If SrcI's central source is a binary, and the measured offset of $\sim5$ AU
between the disk midpoint and the central emission source is real, we can guess
that the binary's orbit is $\lesssim5$ AU.  For such an orbital radius, the
orbital timescale is only $\sim3$ years.   It will therefore be productive to
re-observe \sourcei over the next several years to see if the hot spot moves on
such a timescale.

\section{Conclusions}
\label{sec:conclusions}
We report observations that resolve \sourcei's disk in both continuum
and line emission.  We measure the mass of \sourcei by fitting the 
rotation curve with a Keplerian disk model, finding the following:

\begin{enumerate}
    \item The central source has mass $M=15\pm2$ \msun, where the the error
        bar represents the range of consistent models rather than a typical
        $1-\sigma$ statistical uncertainty. 
    \item The \water $5_{5,0}-6_{4,3}$ line is not masing and kinematically
        traces both the upper envelope of the disk and the lower portion of the
        outflow. \\
    \item We observe several lines that trace the disk kinematics
        directly, though the molecules producing these lines remain
        unidentified.  These lines are visible only toward the outskirts of the
        disk and are morphologically distinct from both the \water and SiO
        lines that follow the outflow. 
    \item A compact source in the approximate center of the disk
        is resolved at 1.3 mm, and it is slightly off-center.  It therefore is
        most likely a hot region of the inner disk.  It may be produced
        by time-varying illumination from an unequal mass binary.
\end{enumerate}

The mass we have measured is higher than in several recent publications because
both the resolution and sensivity of our observations were greater.  These new
data allowed us to identify and measure the spectral features that directly
trace the disk kinematics, while previous data convolved the disk and outflow
kinematics.  This higher measured mass implies that the dynamical decay
scenario for the \sourcei - BN - \sourcex system is viable.

\acknowledgements{
We thank the anonymous referee for a thorough and helpful review.
This paper makes use of the following ALMA data: ADS/JAO.ALMA\#2016.1.00165.S
ALMA is a partnership of ESO (representing its member states), NSF (USA) and
NINS (Japan), together with NRC (Canada), MOST and ASIAA (Taiwan), and KASI
(Republic of Korea), in cooperation with the Republic of Chile. The Joint ALMA
Observatory is operated by ESO, AUI/NRAO and NAOJ.  The National Radio
Astronomy Observatory is a facility of the National Science Foundation operated
under cooperative agreement by Associated Universities, Inc.}

\software{
The software used to make this version of the paper is available from github at
\url{https://github.com/keflavich/Orion_ALMA_2016.1.00165.S}
(\url{https://doi.org/10.5281/zenodo.1181877}) with hash \githash
(\gitdate).  The tools used include \texttt{spectral-cube}
(\url{https://doi.org/10.5281/zenodo.591639} and
\url{https://github.com/radio-astro-tools/spectral-cube})
and
\texttt{radio-beam} (\url{https://github.com/radio-astro-tools/radio-beam},
\url{https://doi.org/10.5281/zenodo.1181879}) from the
\texttt{radio-astro-tools} package
( \url{radio-astro-tools.github.io}), \texttt{astropy}
\citep{Astropy-Collaboration2013a}, \texttt{astroquery}
(\url{astroquery.readthedocs.io}, \url{https://doi.org/10.5281/zenodo.591669} )
and \texttt{CASA} \citep{McMullin2007a}.
}

\appendix

\section{Continuum Modeling Figures}
\label{appendix:contmodel}
In this appendix, we show figures illustrating the continuum modeling process.
Figures \ref{fig:contmodel_residuals_B6} and \ref{fig:contmodel_residuals_B3}
show the model and residuals for the band 6 and band 3 data, highlighting the
significantly improved fit as more model parameters are added.

In Figure \ref{fig:contmodel_residuals_B6}, the apparent point sources at the
edge of the disk in the central column are artificial features introduced by
the model; since the model is forced to be smooth, the best-fit model is one
that is more centrally peaked, which results in an under-prediction of the disk
brightness toward the edges.  

Also in this figure, there appears to be a faint `halo' of emission at the
$\sim30$ K level around the modeled region.  The halo is asymmetric, with a
greater extent toward the southwest.  If this feature is not an artifact of the
data reduction, which we cannot rule out, it is likely to be from optically
thin dust above and below the disk, since it is not detected in the 3.2 mm data.

\Figure{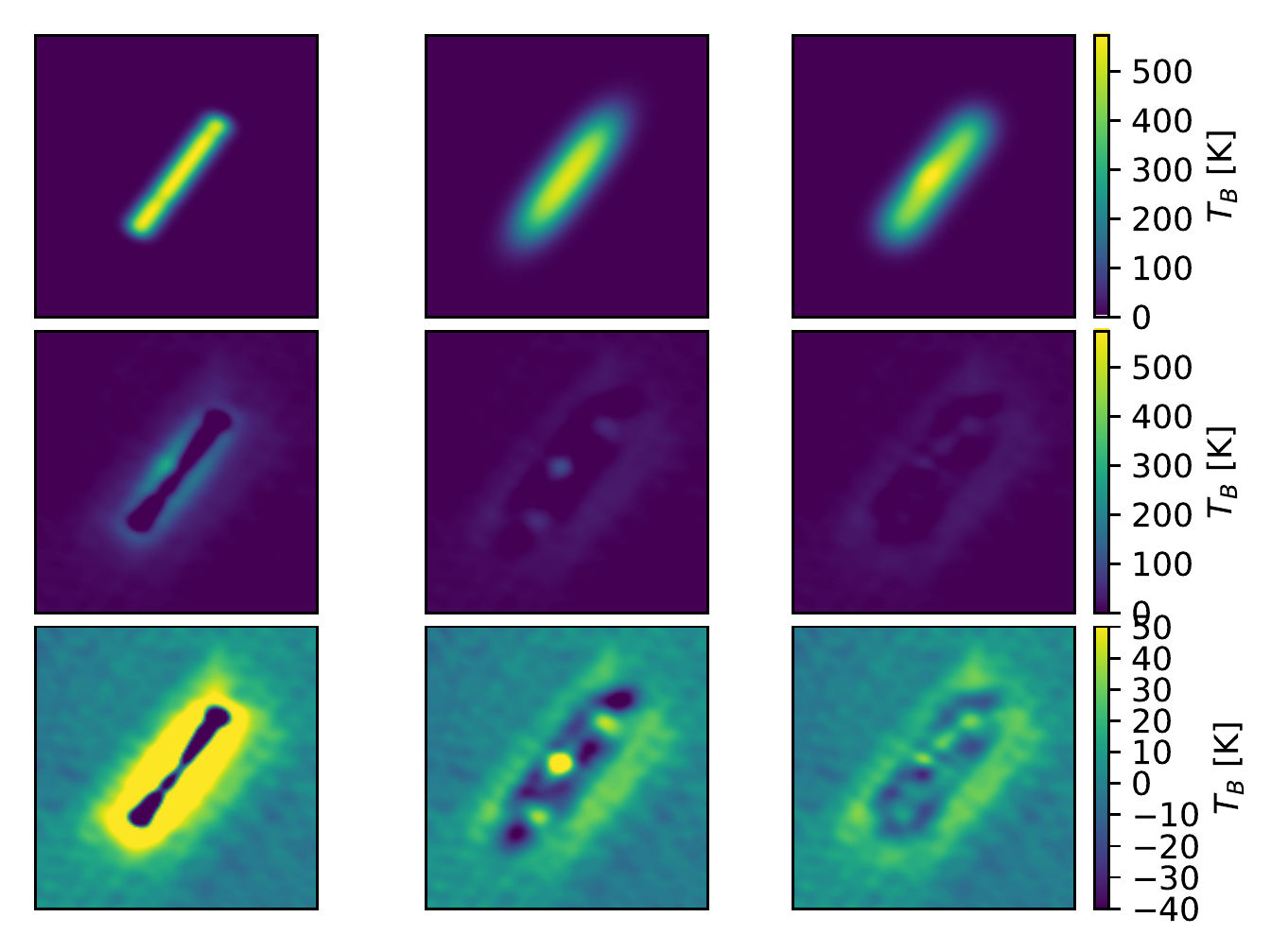}
{A series of plots showing the band 6 continuum models used and their residuals.
The top row shows the models, starting from a simple 1D linear model convolved
with the beam (left), continuing with a disk smoothed with a broader beam to
account for scale height (middle), and finally a version of the middle model
with a smeared point source added (right).  The fit parameters are given in Table
\ref{tab:continuum_fit_parameters}.  The second and third row show the
residuals (data - model) for each of the models in the top row; the bottom row
uses a narrow linear scale to emphasize the lower-amplitude residuals, while
the top two use an arcsinh stretch to display the full dynamic range.
}
{fig:contmodel_residuals_B6}{1}{7.5in}

\Figure{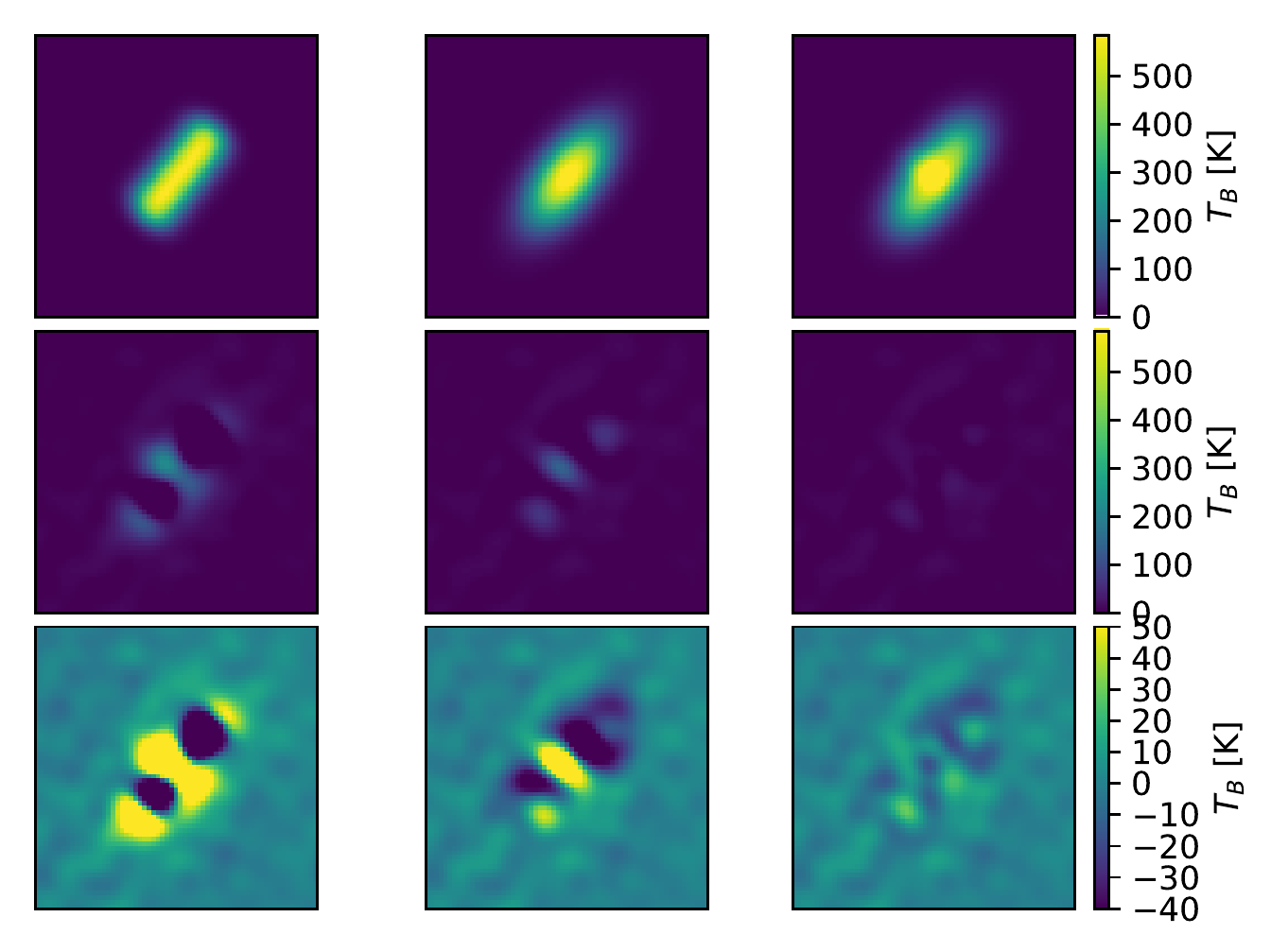}
{A series of plots showing the band 3 continuum models used and their residuals.
See the caption of Figure \ref{fig:contmodel_residuals_B6} for details.
}
{fig:contmodel_residuals_B3}{1}{7.5in}

\section{The SiO outflow in PV space}
\label{sec:siopv}
To illustrate the change in velocity structure with height from the disk, 
we show position-velocity diagrams of SiO v=0 J=5-4 in Figure \ref{fig:siopv}
and $^{29}$SiO J=5-4 in Figure \ref{fig:29siopv}.
These images are extracted from equal distances above and below
the disk midplane.  At greater distances from the midplane, the high-velocity,
low-separation features fade out, while more low-velocity material
becomes visible at larger separations.  These structures are similar to what was
shown in the 484 GHz Si$^{18}$O J=12-11 line in Figure 2 of
\citet[][]{Hirota2017b}.  Our data are consistent with their interpretation
that the SiO isotopologues trace a rotating, expanding outflow.  In the
innermost slice, which shows the SiO emission that just skirts the edges of the
disk, the velocity curve is consistent with the 15 \msun Keplerian curve
overlaid.

\FigureFour
{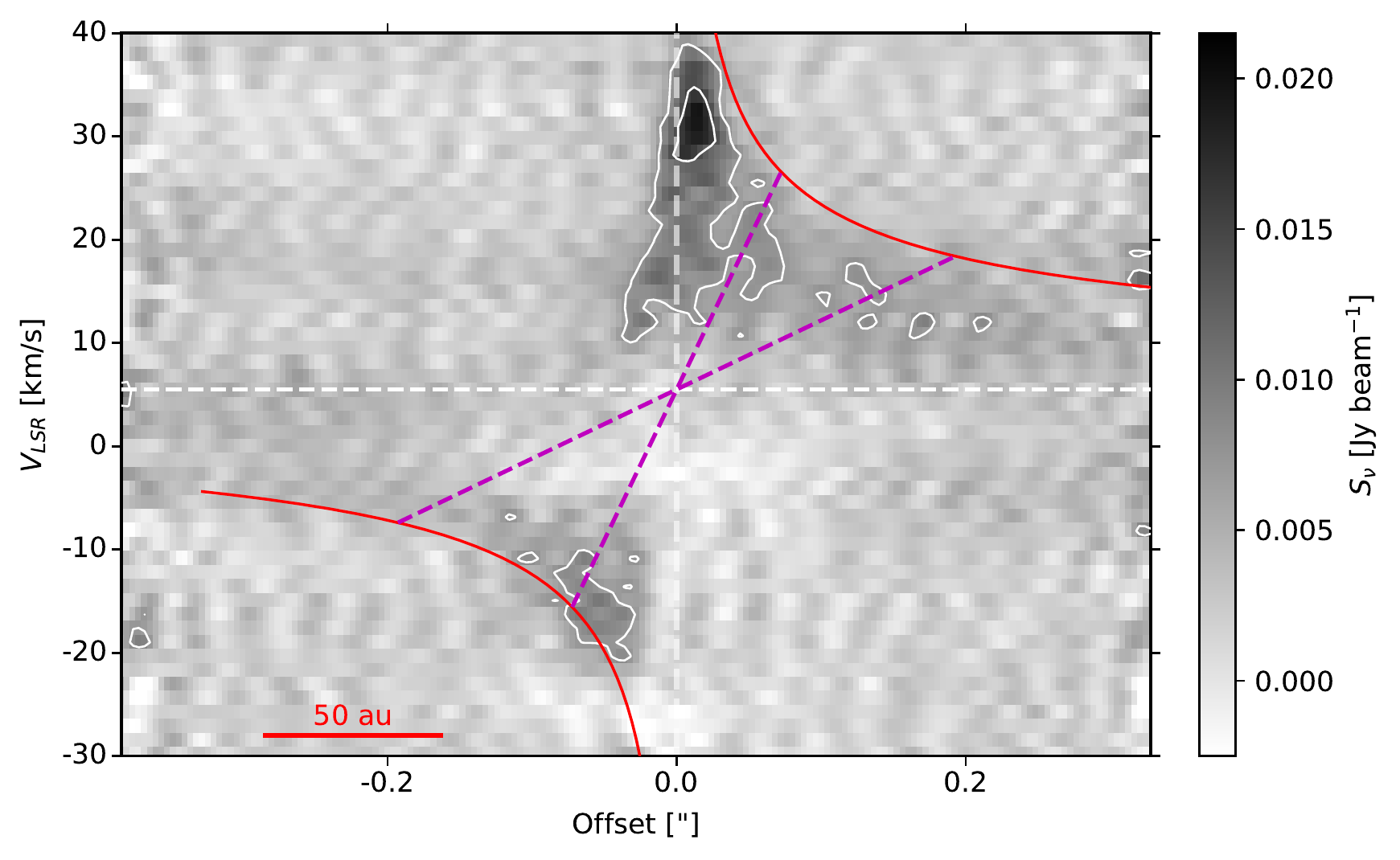}
{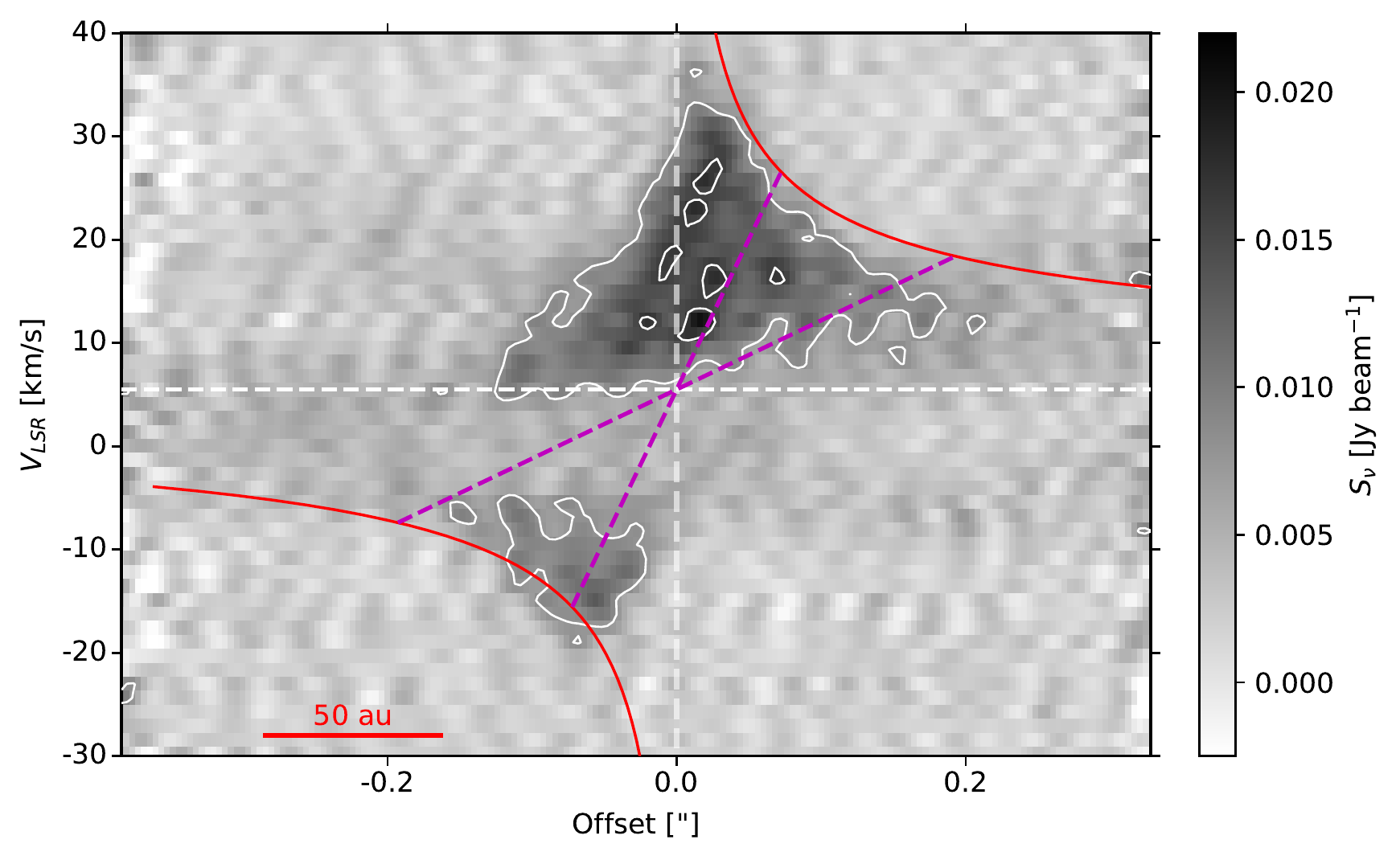}
{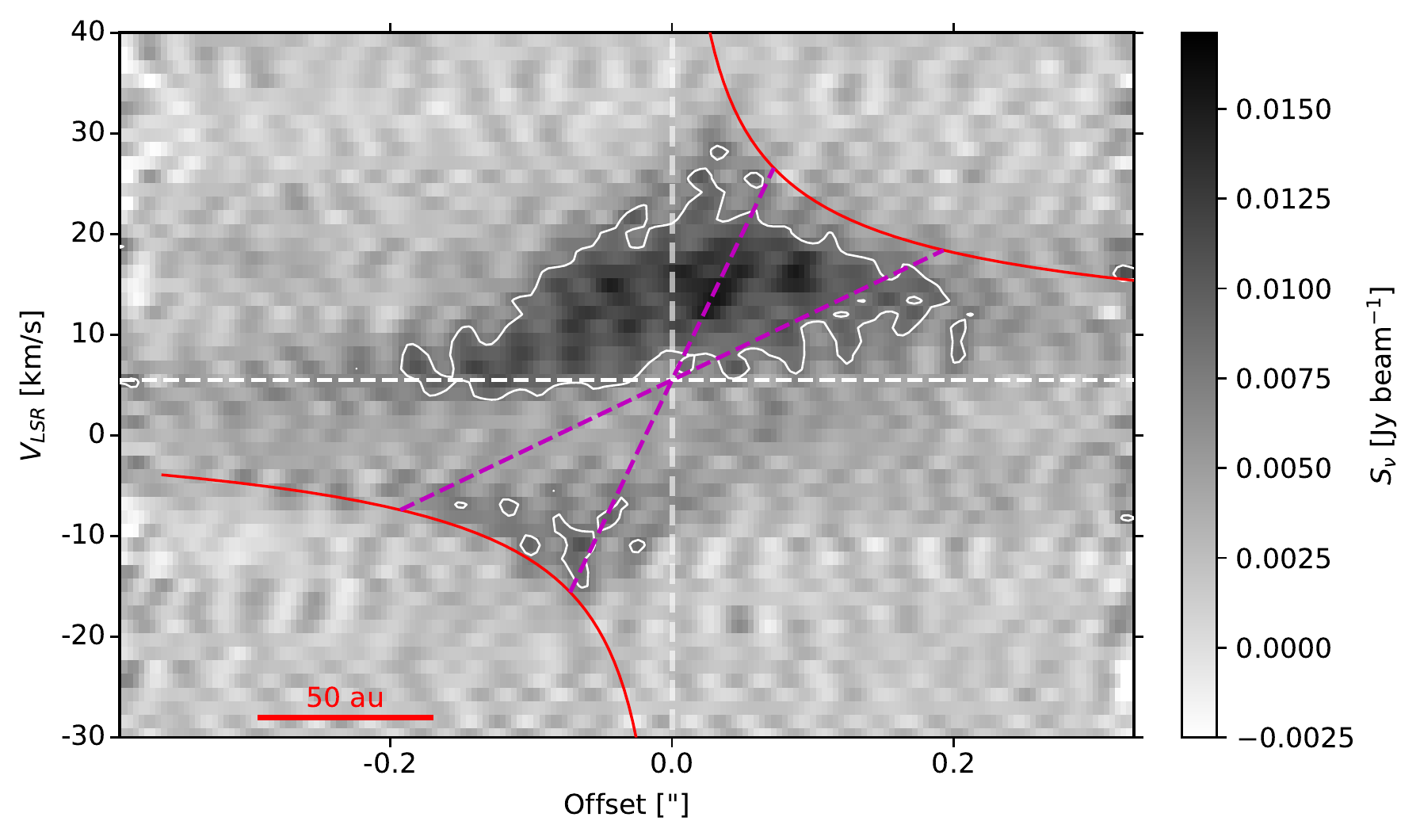}
{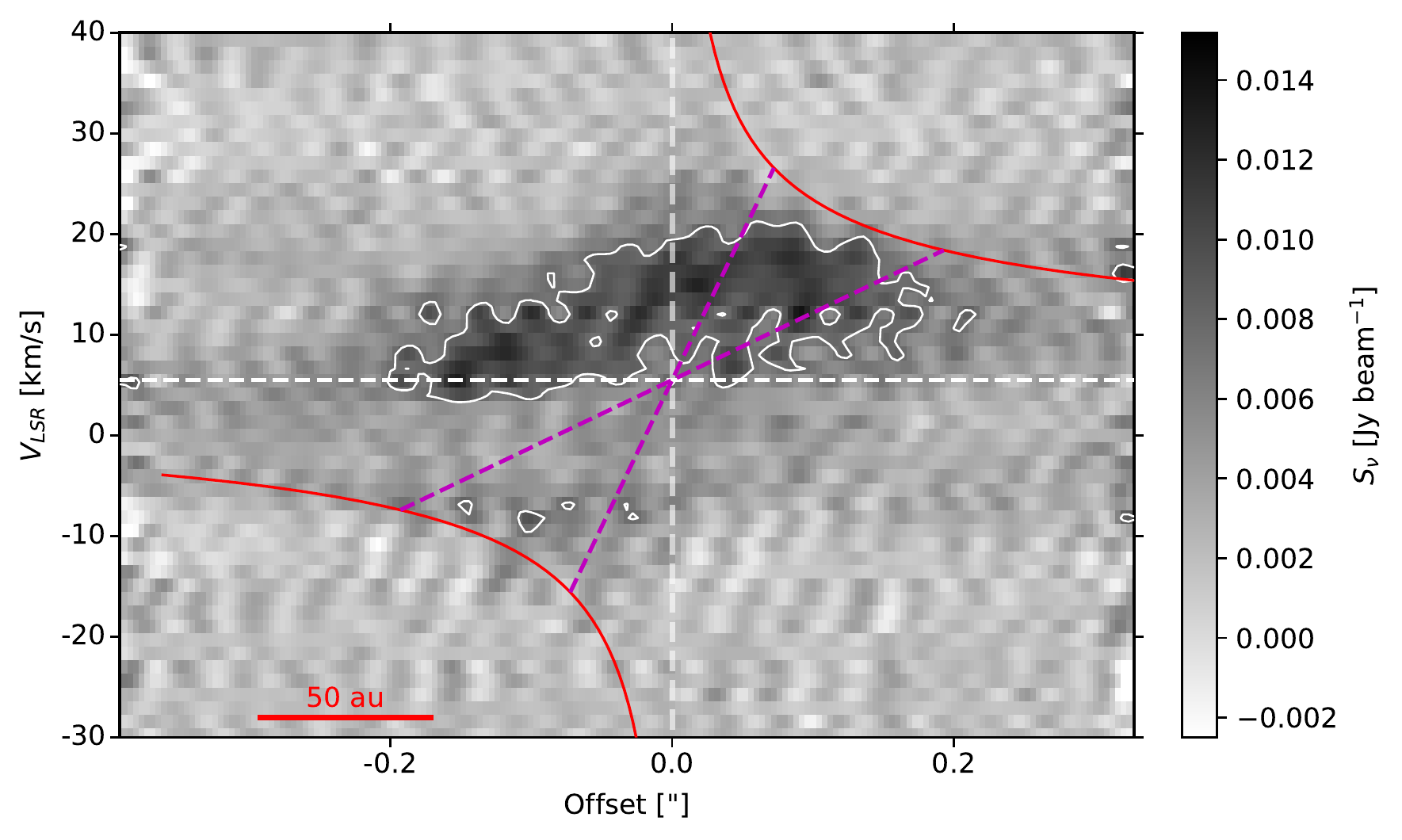}
{Position-velocity slices of SiO v=0 J=5-4 along the disk direction at four heights:
(a) $|h|<0.05\arcsec$,
(b) $0.05\arcsec<|h|<0.1\arcsec$,
(c) $0.1\arcsec<|h|<0.15\arcsec$,
(d) $0.15\arcsec<|h|<0.2\arcsec$.
Contours are overlaid at 5 and 10 $\sigma$.
These images are produced from the robust -2 weighted cubes.
The missing emission around $v=0$ \kms is likely caused by image filtering
effects; at these velocities, there is extended, smooth SiO emission from the
surrounding cloud.
}
{fig:siopv}{1}{0.23\textwidth}

\FigureFour
{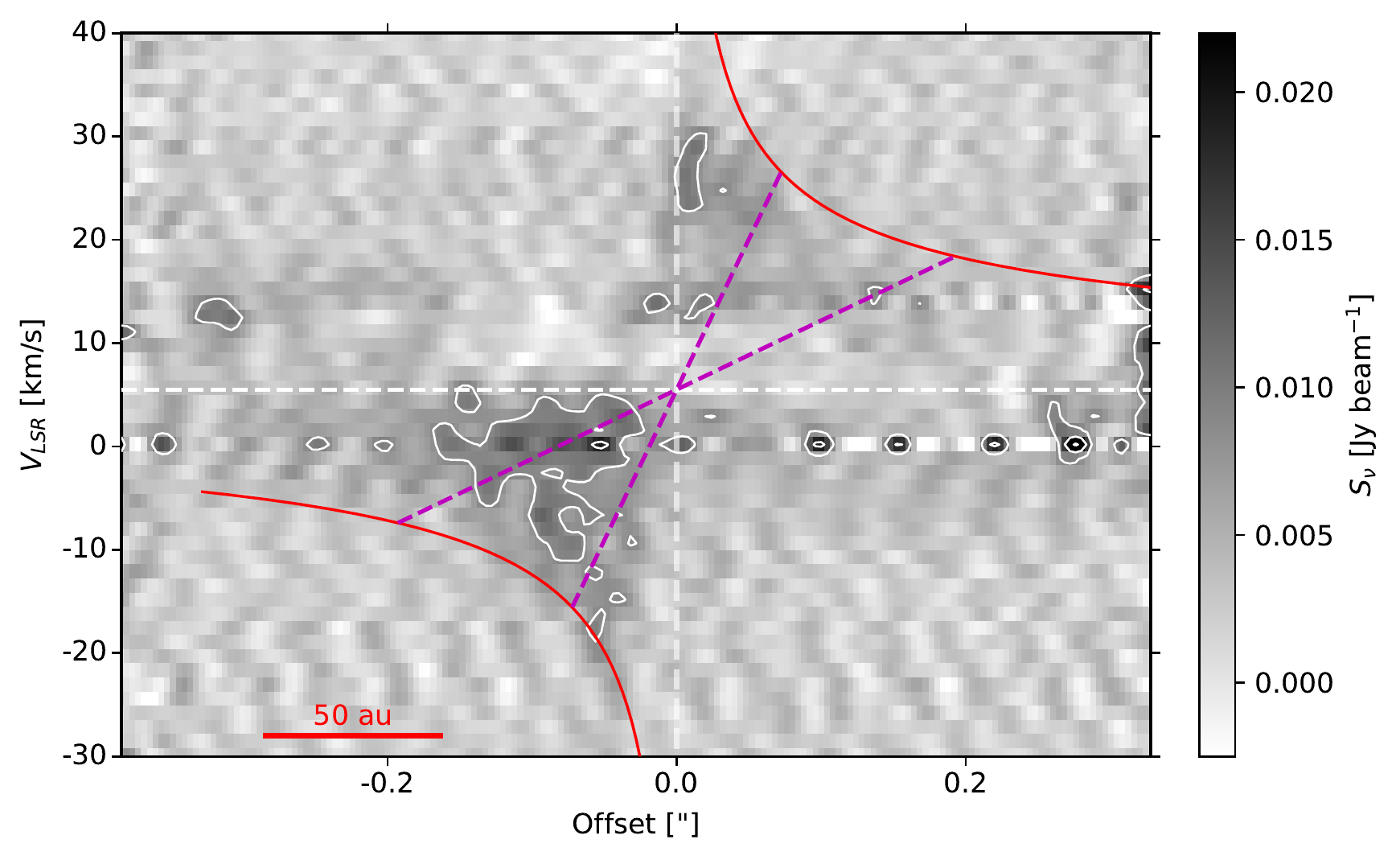}
{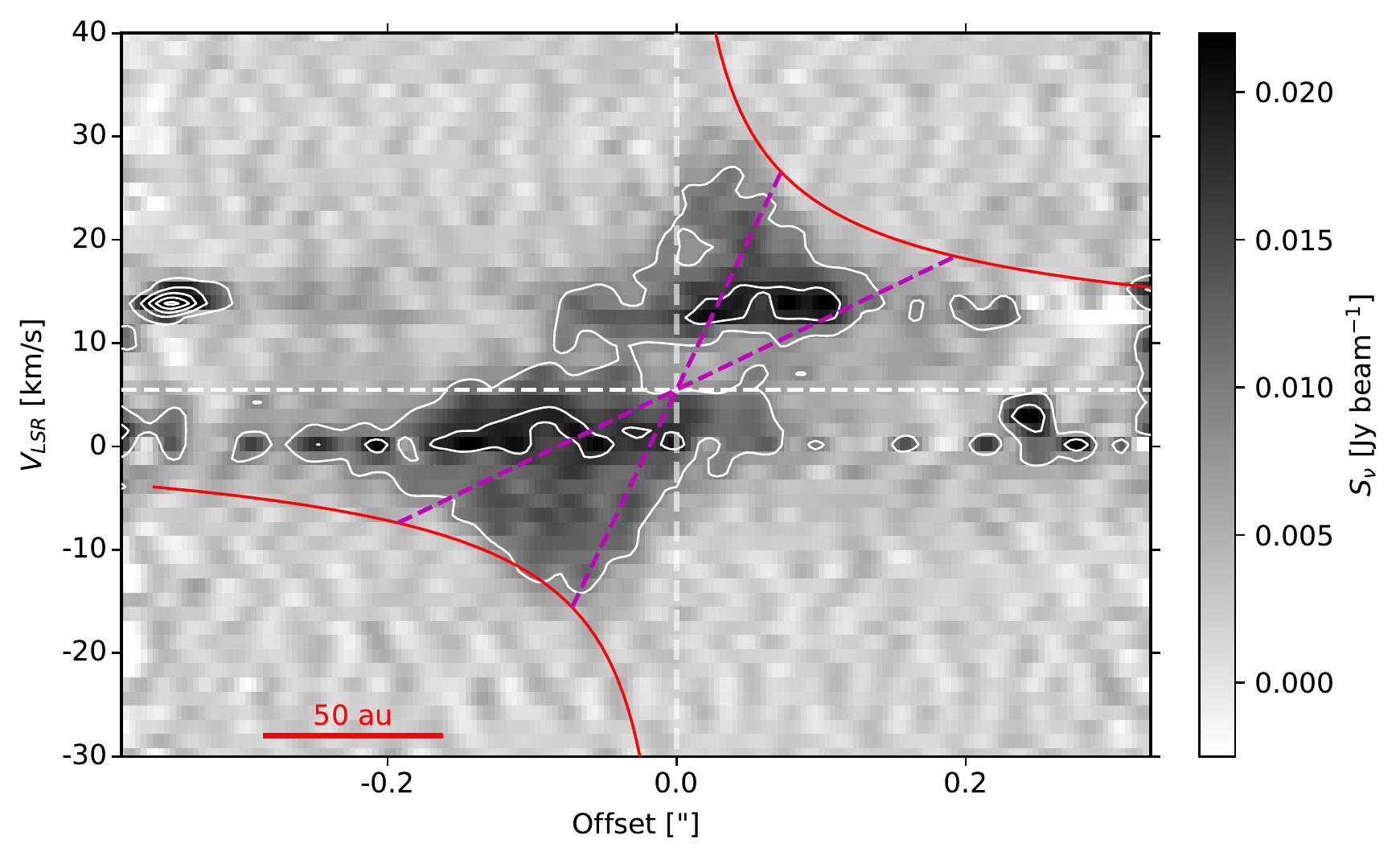}
{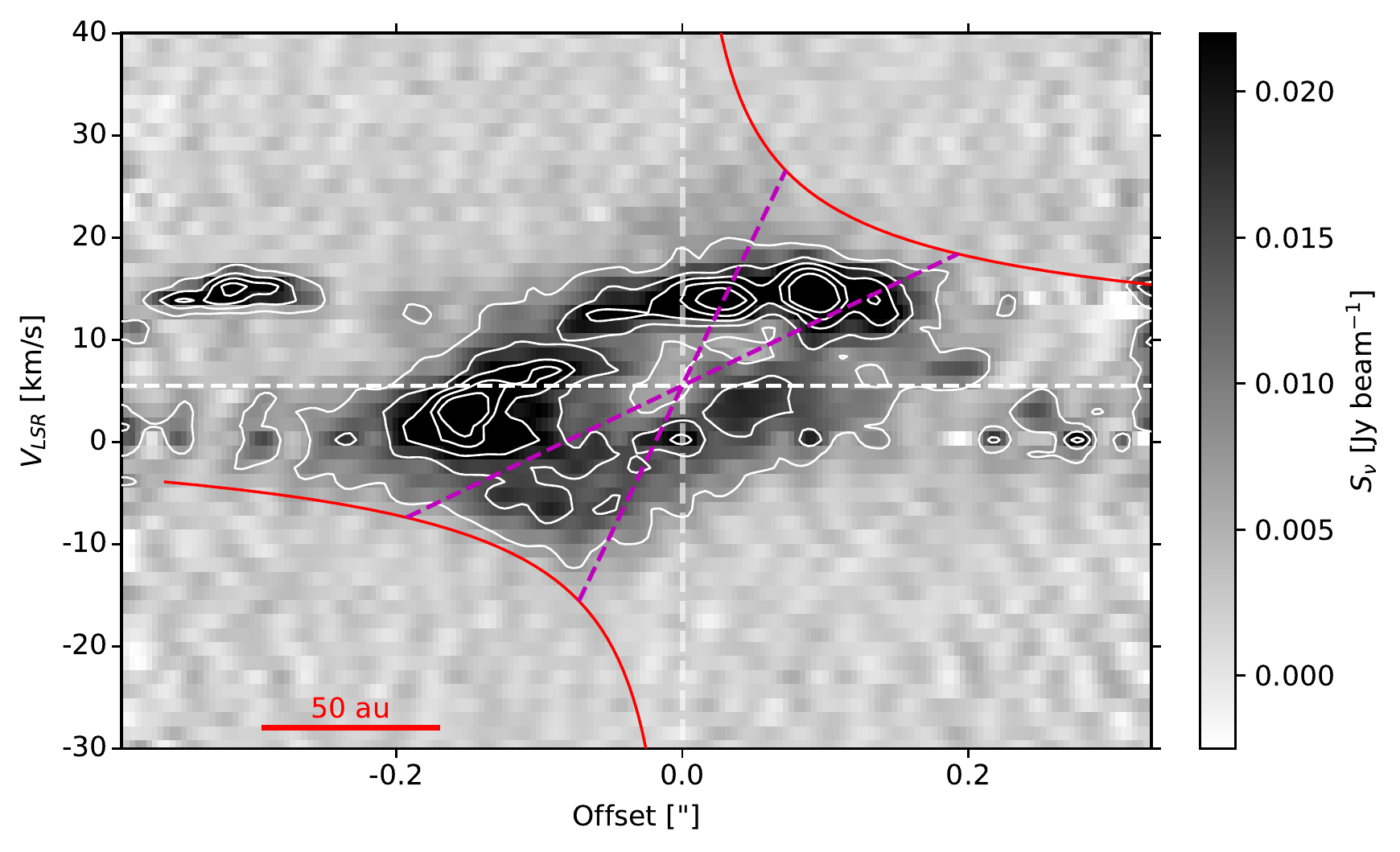}
{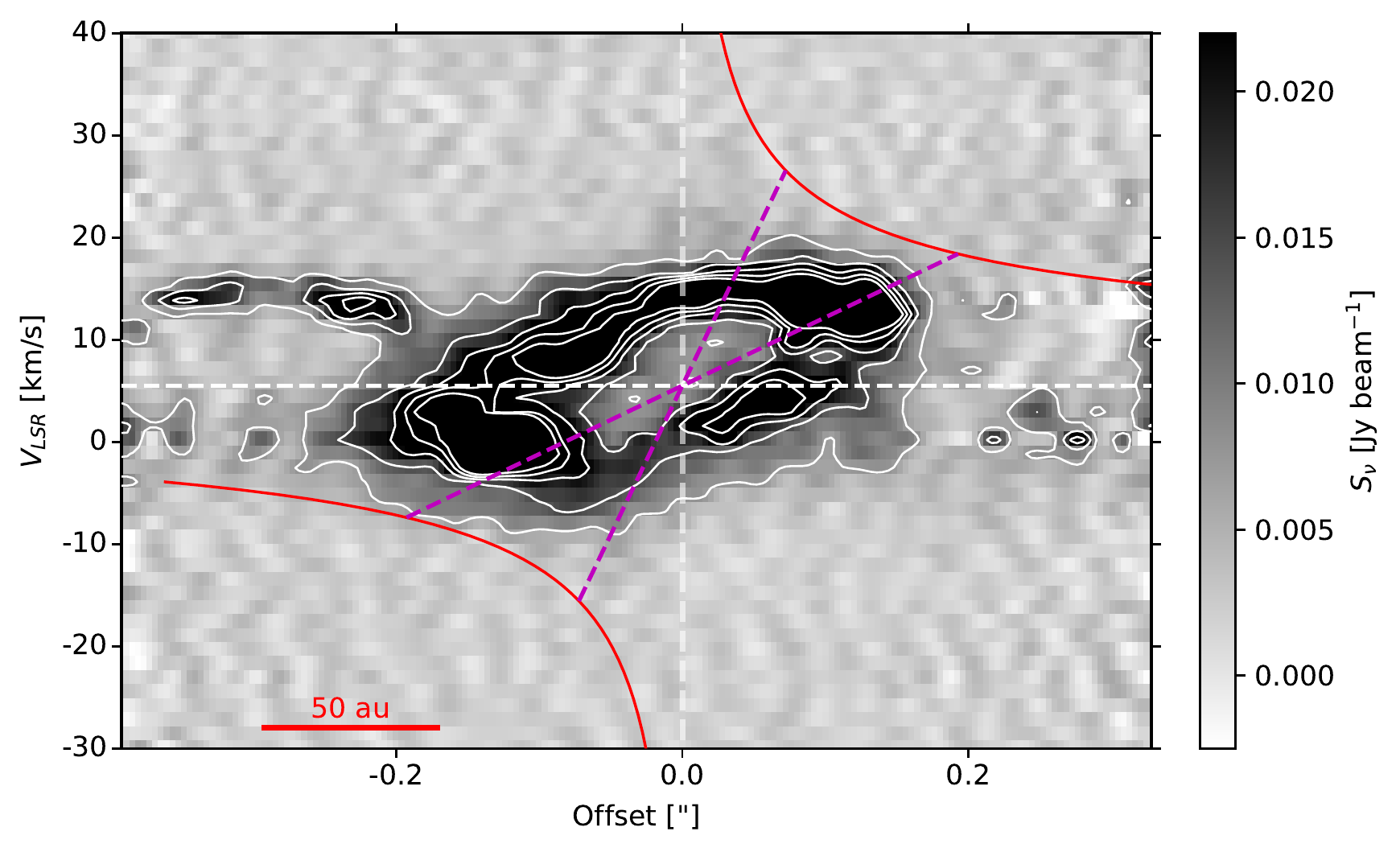}
{Position-velocity slices of $^{29}$SiO v=0 J=5-4 along the disk direction at four heights:
(a) $|h|<0.05\arcsec$,
(b) $0.05\arcsec<|h|<0.1\arcsec$,
(c) $0.1\arcsec<|h|<0.15\arcsec$,
(d) $0.15\arcsec<|h|<0.2\arcsec$.
Contours are overlaid at 5, 10, 15, 20, and 25 $\sigma$.
These images are produced from the robust -2 weighted cubes.
While similar to the $^{28}$SiO shown in Figure \ref{fig:siopv},
there is a remarkable position-velocity ring at high elevations
that is coincident with many of the SiO and \water masers.
}
{fig:29siopv}{1}{0.23\textwidth}

\section{A deeper examination of the water line: Evidence that it traces the disk kinematics}
\label{appendix:waterlinerevisited}
The \water-derived mass presented in Section \ref{sec:kinematics} relies on the
\water line tracing the disk kinematics.  Since the \water clearly also
traces the outflow, showing the same X-shaped morphology as the SiO, it does
not trace just the disk.

Nonetheless, the midplane position-velocity slice of the \water line does
appear to genuinely trace disk kinematics.  Qualitatively, the PV diagram
appears exactly as expected for a disk with an inner and out radial cutoff.

To assess possible contamination from the outflow, we compare position velocity
slices at different vertical displacements from the disk center in Figure
\ref{fig:waterpvheights}.
The left panel shows the kinematic signature we attribute to the disk, which
closely resembles that predicted for a pure Keplerian rotation curve.
In contrast, the middle panel is likely
dominated by outflow emission, since it shows material 0.05-0.1\arcsec (20-40
AU) above the disk, i.e., just outside the 1-$\sigma$ height of the continuum
disk.  
While the outflow continues to show some motion similar to that of the disk, it
lacks the characteristic convex shape of a Keplerian orbit at higher velocities
and separation.
Finally, the rightmost panel shows that the water emission nearly disappears
at heights $h>0.1\arcsec=40$ AU while the maximum velocities observed get
smaller ($dv < 10$ \kms), suggesting that rotation slows in the outflow.

Figure \ref{fig:waterpvheights} also characterizes some of the `forbidden' velocity
components, i.e., those seen in quadrants 2 and 4.  These components get stronger
at higher vertical positions on the disk, implying that they come from the outflow,
not the disk.
The ``ring'' shape observed in the high-latitude figures indicates the outflow
is expanding \citep[see, e.g., the model in supplementary figure 1
of][]{Hirota2017b}.
The velocity asymmetry, which shows an excess toward the red side of the disk
and outflow, is also present in SiO.  We do not have a straightforward
explanation for this asymmetry except to assert that it implies an
asymmetry in the direction of mass ejection in the outflow.
These velocity components are unlikely to be produced by infall motions, since
they are observed perpendicular to the disk along the direction of the outflow.

\FigureThree
{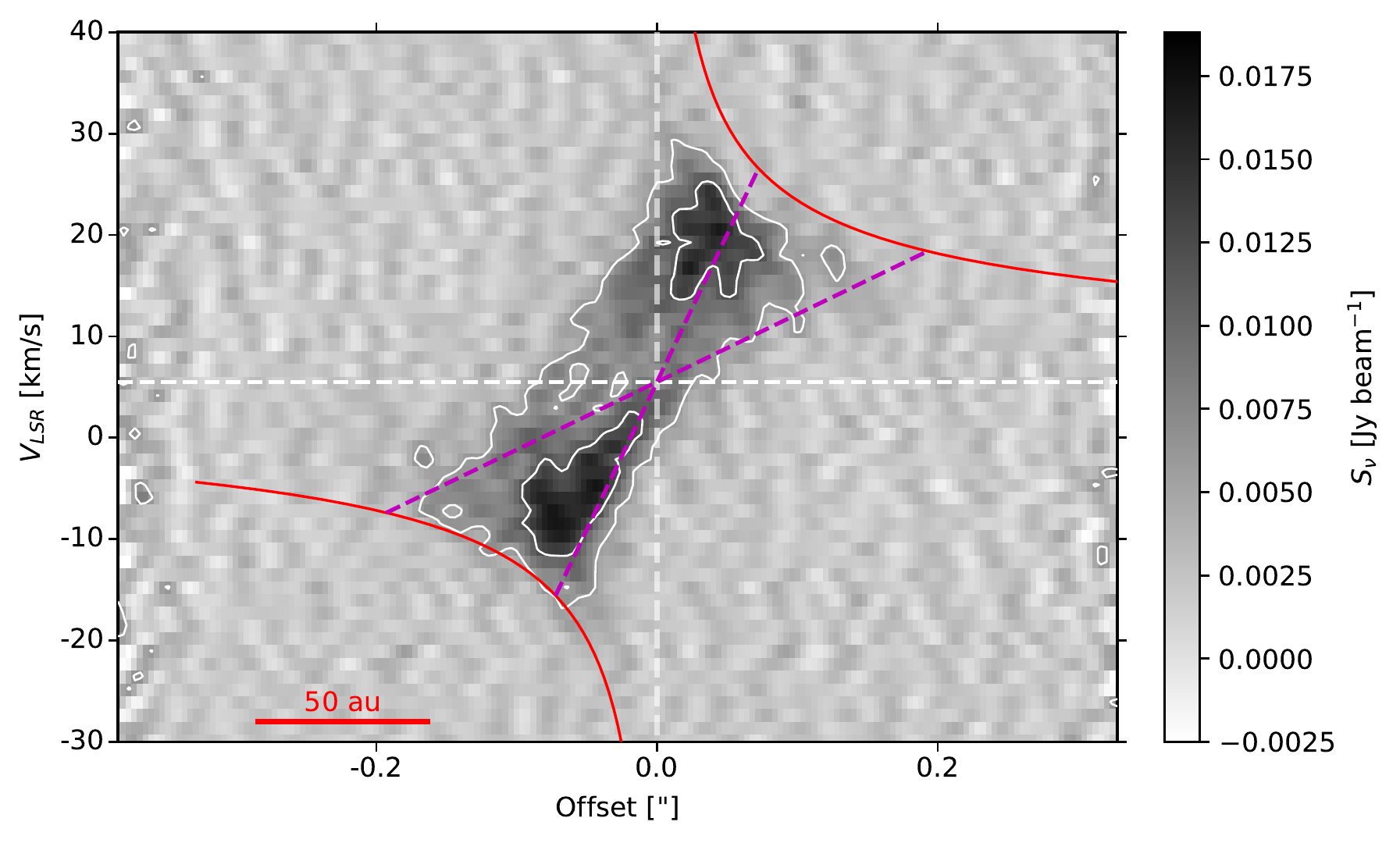}
{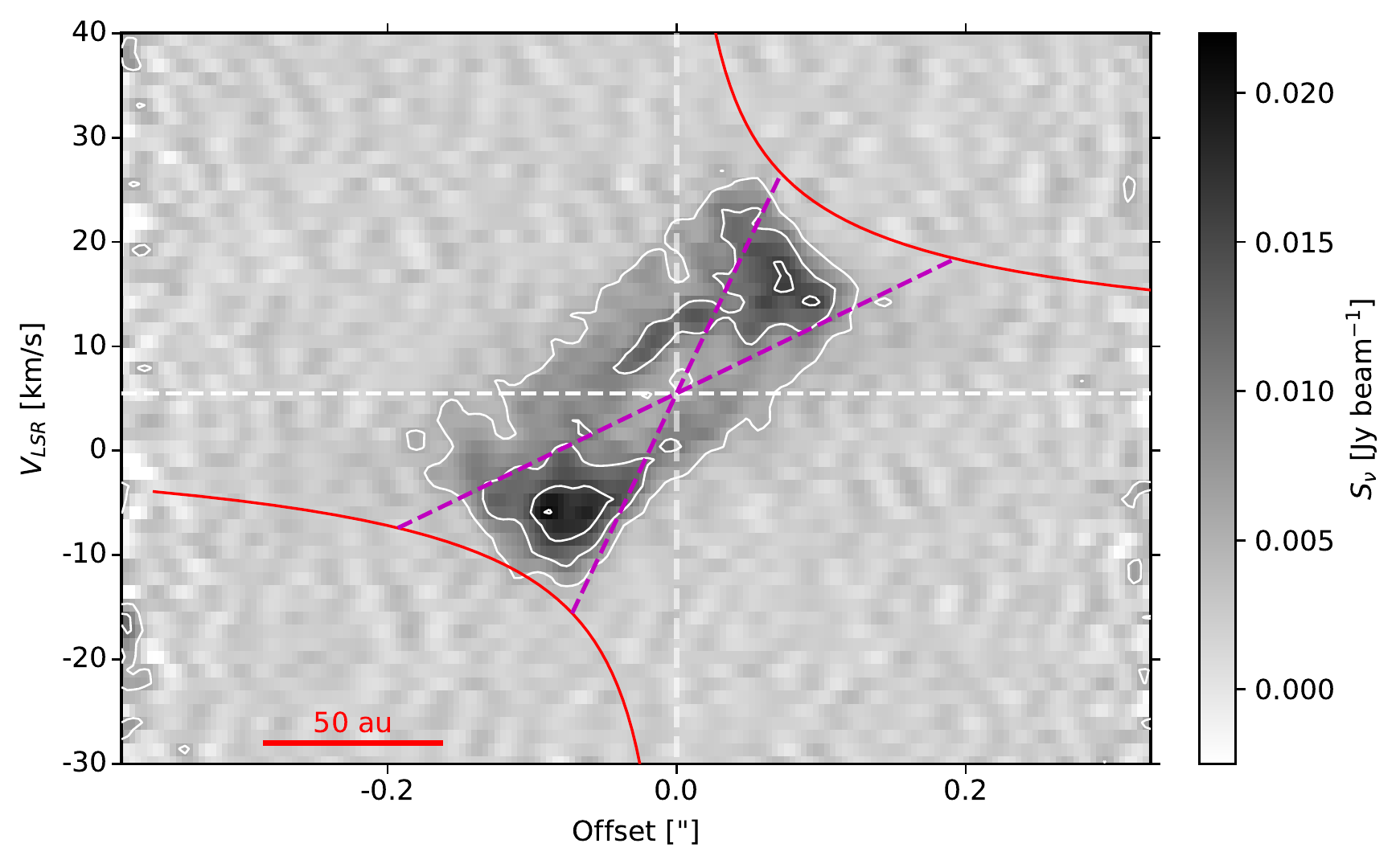}
{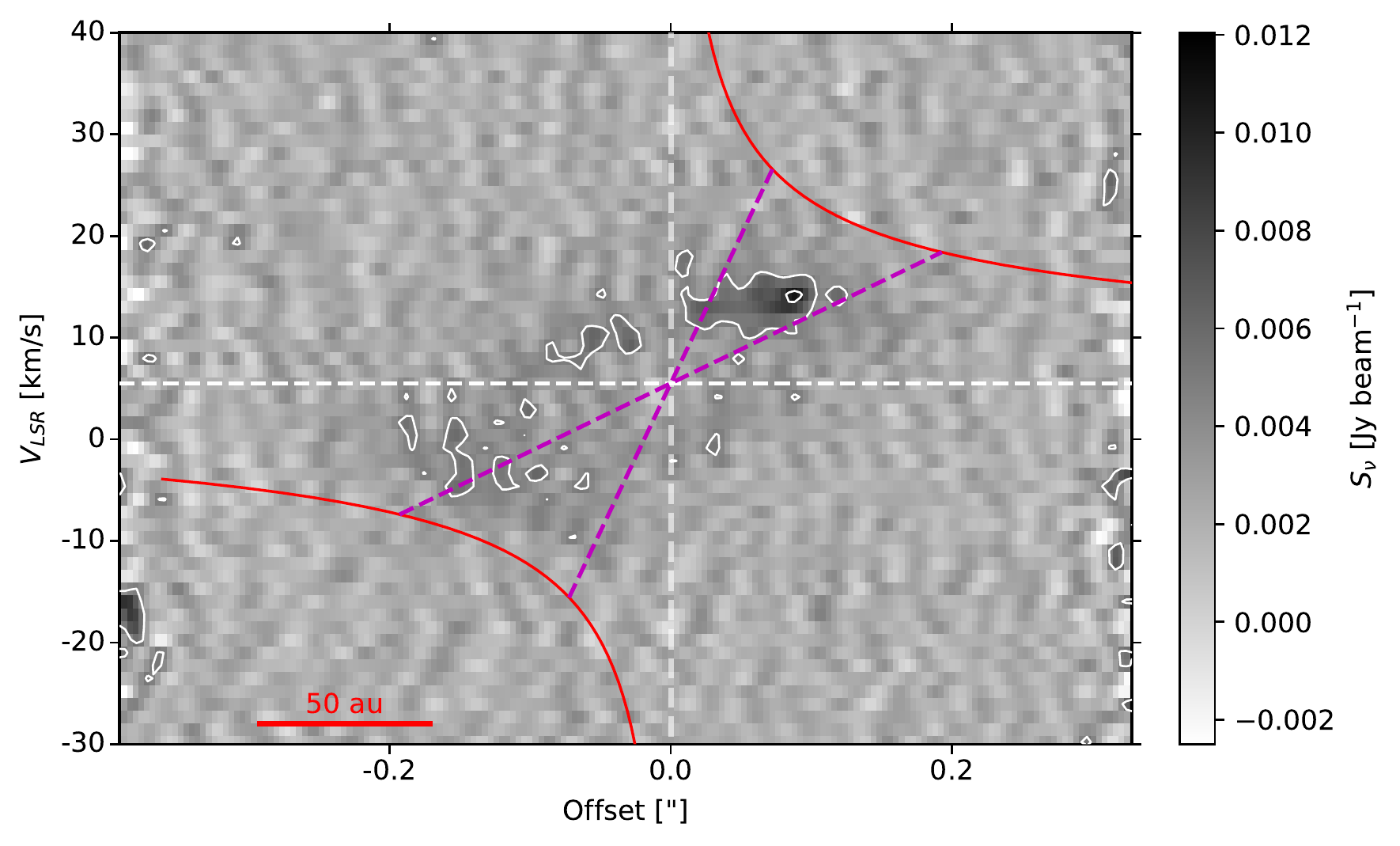}
{Position-velocity slices of the \water $5_{5,0}-6_{4,3}$ line at different
heights above the disk plane from the robust -2 data cube.  The left panel
shows the inner 0.1\arcsec (i.e., near the midplane, $|h| < 0.05\arcsec$,
$|h|<20$ AU), the middle shows the range $0.05\arcsec < |h| < 0.1$\arcsec, and
the right shows the range $0.1\arcsec < |h| < 0.15\arcsec$.  All three panels
show averages over the specified range, with the colorbar showing intensity in
mJy \perbeam.
Contours are overlaid at 5, 10, 15, and 20 $\sigma$.
The red solid lines show the Keplerian
profile for a 15 \msun
central source, and the purple dashed lines show the orbital
track for a particle at 30 and 80 AU for such a source; these are included
primarily to guide the eye.  The middle and right panel are dominated by the
outflow, while the left panel is dominated by the Keplerian orbital profile.
}
{fig:waterpvheights}{1}{2.2in}

\FigureTwo
{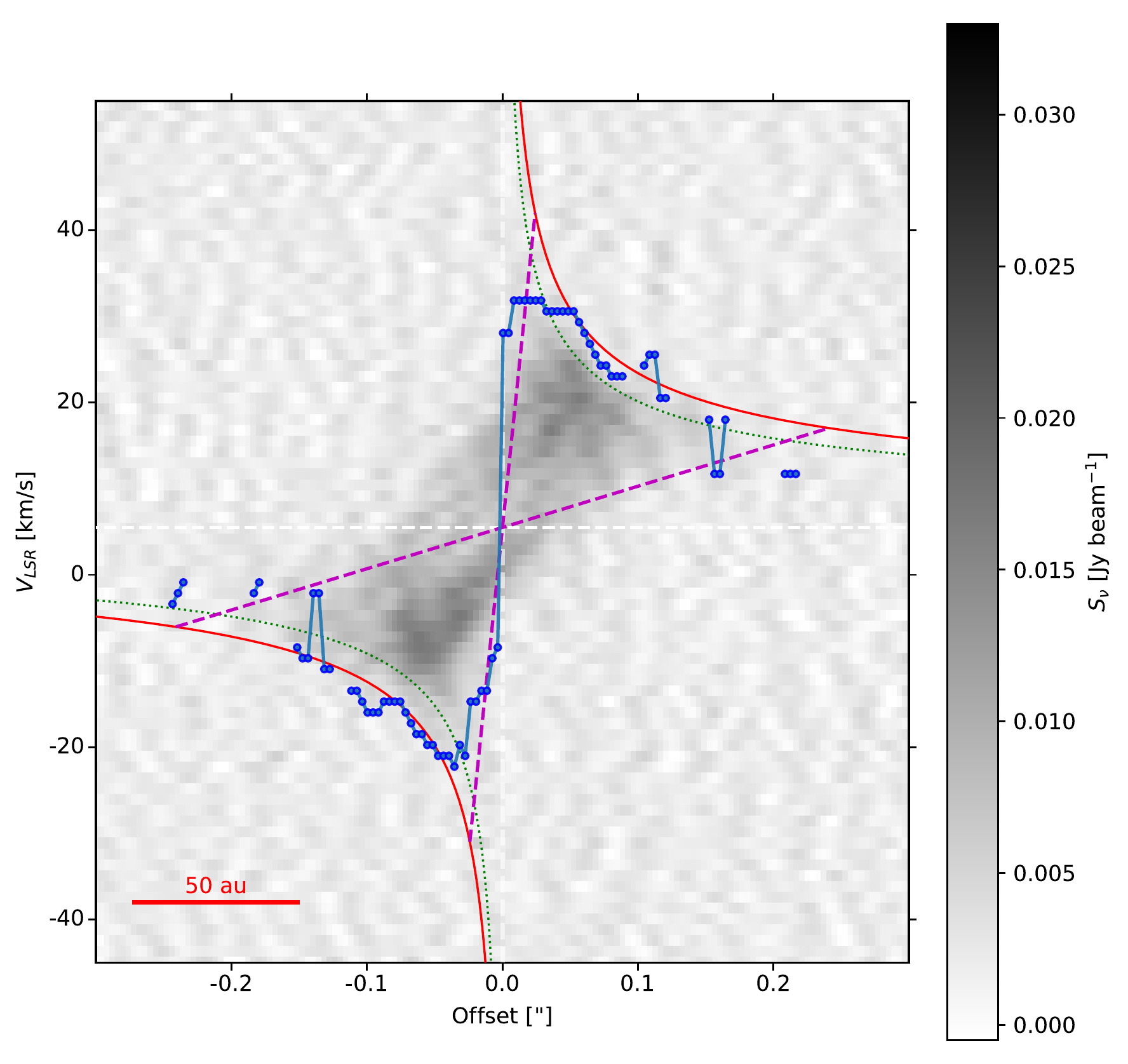}
{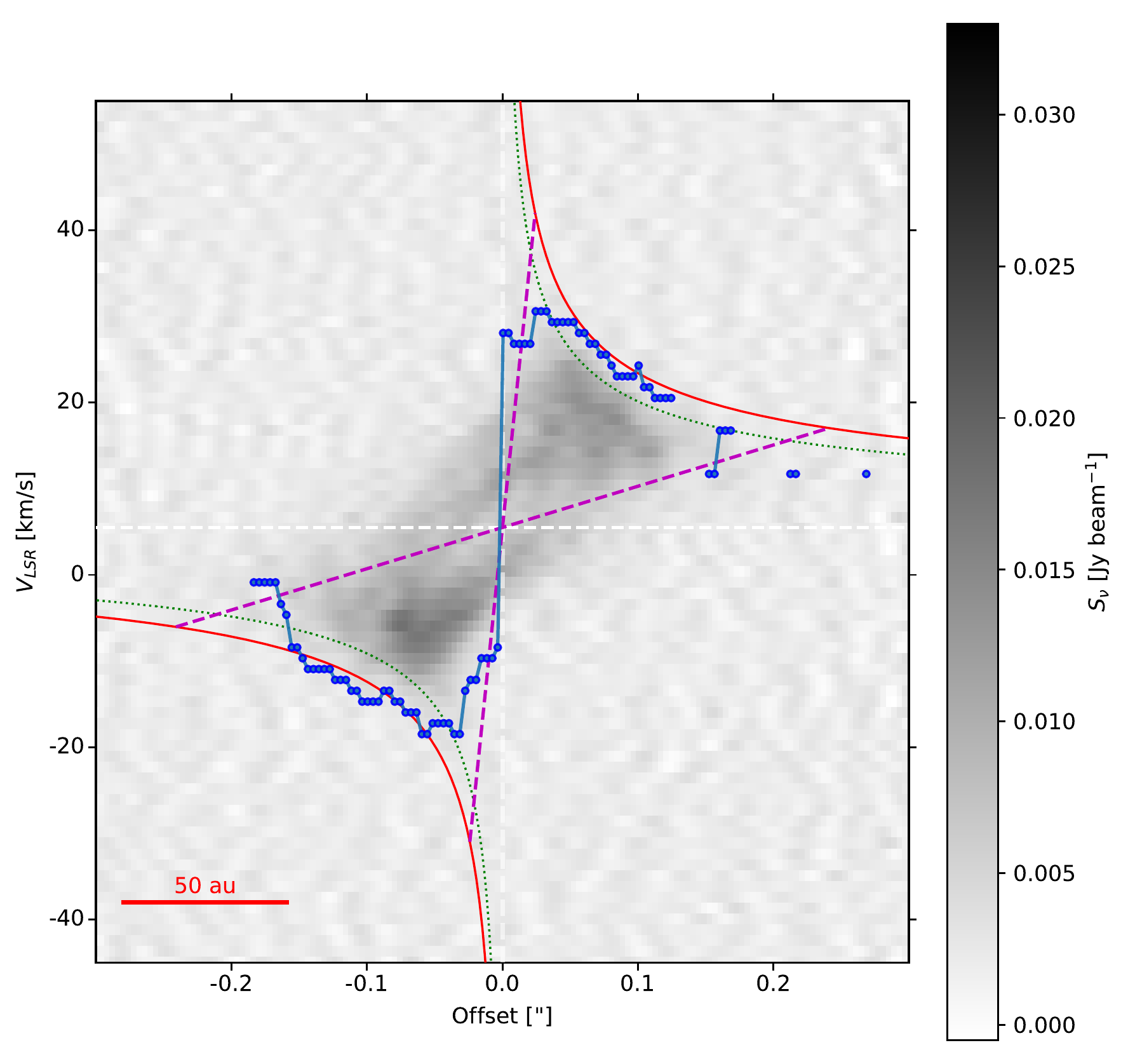}
{
Duplicate of Figure \ref{fig:h2okepler} for robust -2 data with different
vertical extents included in the position-velocity slice.
The left figure shows the average PV slice over the vertical range range
$h=\pm0.05\arcsec$, and the right right shows the same with $h=\pm0.1\arcsec$.
}
{fig:h2okepler_rm2}{1}{3.5in}

\section{Stacked Spectra}
\label{appendix:stackedspectra}
We reported the detection of several unidentified lines.
To measure their frequencies precisely, we performed a stacking
analysis in which we adopt the velocity field of the U232.511 line,
shift all spectra across the disk to the same velocity frame, and average them.
We stacked the robust 0.5 cubes, as the surface brightness sensitivity
of the robust -2 cubes was too poor to justify stacking.
We then fit the lines with Gaussians to determine their centroid frequency.
We searched within a narrow range of velocities ($v_{LSR}=3-8$ \kms)
for known lines in the Splatalogue collection of line catalogs using
\texttt{astroquery}.
While many of the lines have plausible carriers within 1-2 \kms, such as
highly-excited CH$_3$OCHO or variants of SO$_2$, there is no consistent pattern
to the detected lines and no individual species can explain more than a few of
the observed lines.  These disk-averaged spectra are shown in Figure \ref{fig:stackplots_labeled}
with the lines labeled.

We list the line frequencies (which we use as line names), fitted Gaussian
widths, and fitted amplitudes from the stacked spectra in Table
\ref{tab:unknown_line_frequencies}.

\begin{table*}[htp]
\centering
\caption{Unknown Line Frequencies}
\begin{tabular}{cccc}
\label{tab:unknown_line_frequencies}
Line Name & Frequency & Fitted Width & Fitted Amplitude \\
 & $\mathrm{GHz}$ & $\mathrm{km\,s^{-1}}$ & $\mathrm{mJy}$ \\
\hline
U214.549 & 214.549 & 4.1 & 0.7 \\
U214.637 & 214.637 & 3.2 & 0.3 \\
U214.742 & 214.742 & - & - \\
U214.940 & 214.940 & 4.3 & 4.6 \\
U215.009 & 215.009 & 4.6 & 2.7 \\
U217.229 & 217.229 & 2.6 & 1.3 \\
U217.547 & 217.547 & 6.8 & 1.2 \\
U217.666 & 217.666 & 7.2 & 1.1 \\
U217.980 & 217.980 & 5.2 & 5.6 \\
U218.584 & 218.584 & 4.5 & 1.6 \\
U229.247 & 229.247 & 4.8 & 5.9 \\
U229.550 & 229.550 & 8.0 & 1.1 \\
U229.682 & 229.682 & 15.5 & 3.0 \\
U229.819 & 229.819 & 4.7 & 1.8 \\
U230.322 & 230.322 & 4.7 & 3.8 \\
U230.726 & 230.726 & 5.8 & 1.5 \\
U230.780 & 230.780 & 6.7 & 5.1 \\
U230.966 & 230.966 & 10.4 & 1.1 \\
U232.163 & 232.163 & 4.2 & 1.5 \\
U232.511 & 232.511 & 6.7 & 6.2 \\
U232.634 & 232.634 & 7.8 & 0.8 \\
U233.171 & 233.171 & 4.2 & 2.3 \\
U233.608 & 233.608 & 6.8 & 1.5 \\
\hline
\end{tabular}

\par The frequencies listed have a systematic uncertainty of about 2 \kms (1.5 MHz) because they are referenced to the U232.511 line, which has an unknown rest frequency.  The rest frequency used for the U232.511 line was selected to maximize the symmetry of the emission around 5 \kms.  Some lines were detected in only part of the disk and therefore had bad or malformed profiles in the stacked spectrum; these have fits marked with -'s.
\end{table*}

\FigureFourVertical
{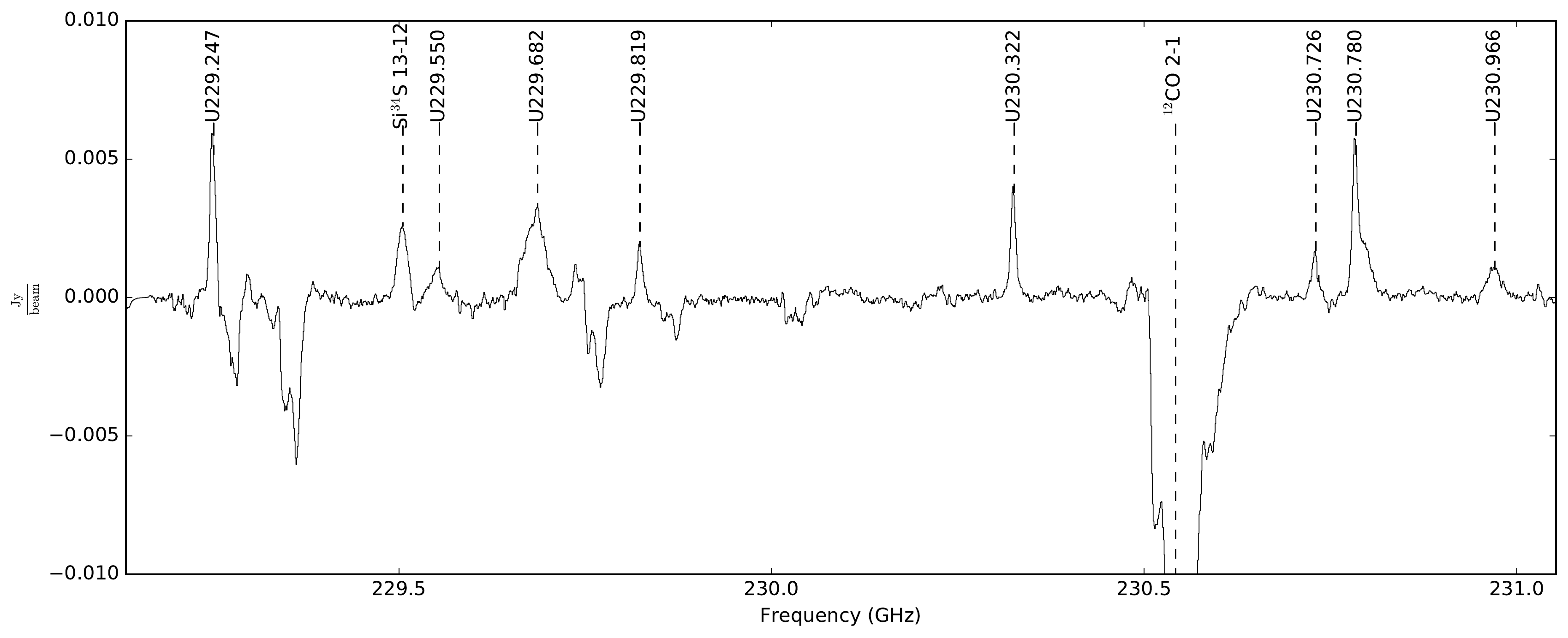}
{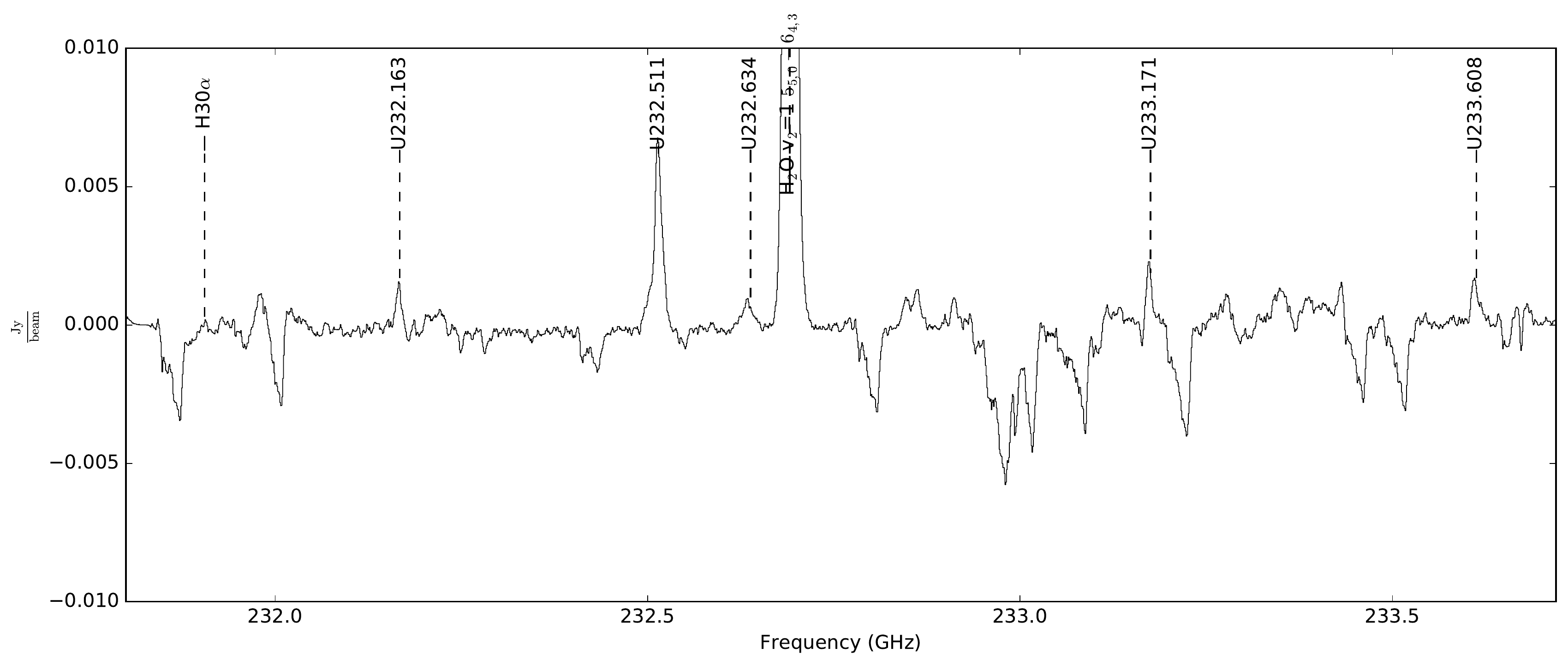}
{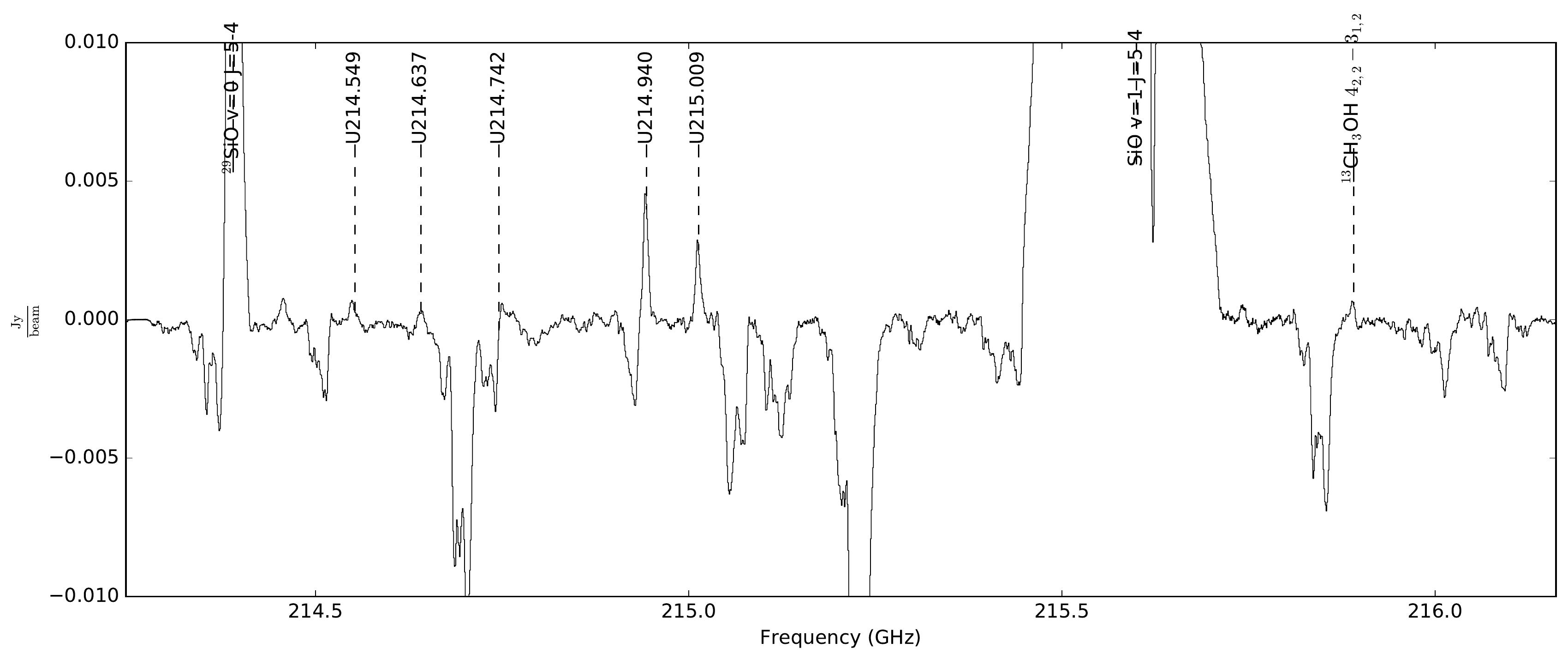}
{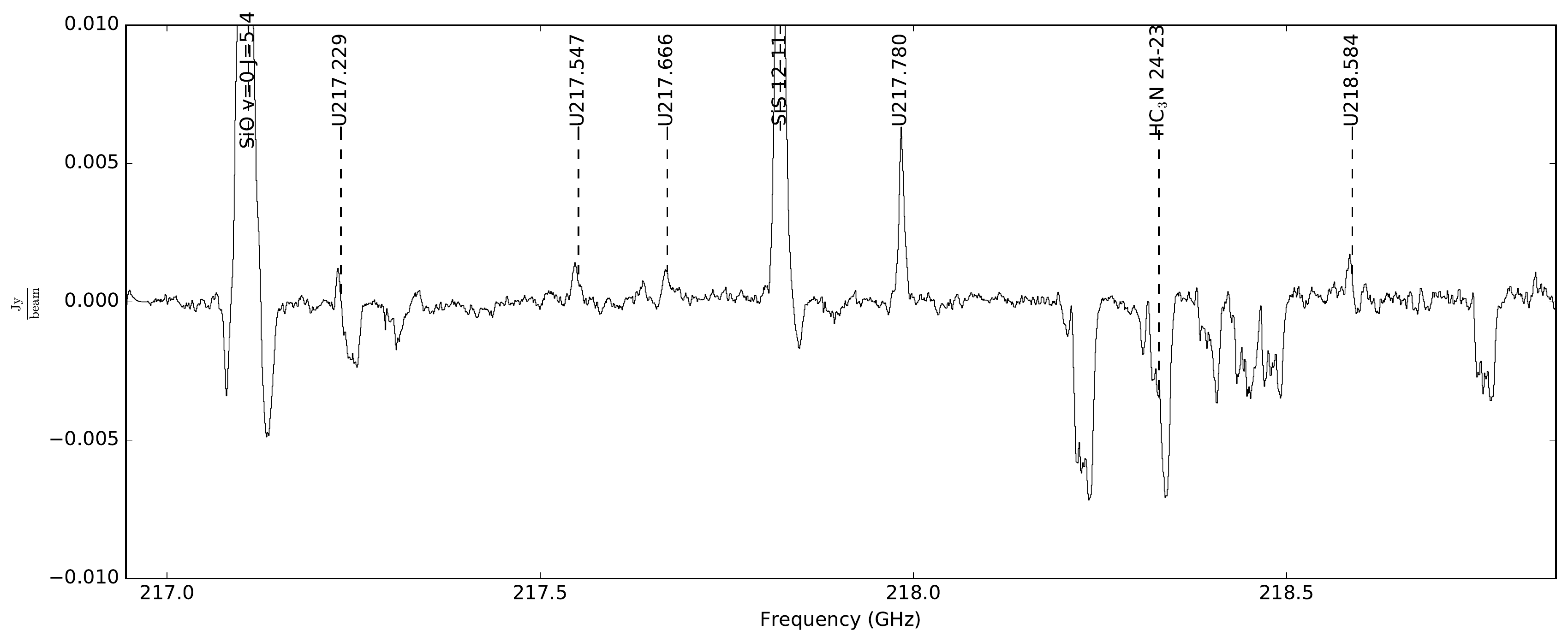}
{Plots of the stacked spectra from spectral windows 0, 1, 2, and 3 (see Table \ref{tab:cube_metadata})
with detected lines labeled.  The spectra are shown with the same y-axis limits; bright SiO and \water
emission is cut off.  In spectral window 2, the region around 215.5-215.6 GHz, near the SiO v=1 J=5-4
maser line (which is the brightest line we detect) is affected by imaging artifacts from the cleaning
process.}
{fig:stackplots_labeled}{0.5}{0.6\linewidth}

\section{Disk parameter determination method comparison}
\label{appendix:centroids}
To compare fairly with \citet{Hirota2014a} and \citet{Plambeck2016a}, we  used
the centroid-velocity method to measure the central source mass.  In this
approach, we fit two-dimensional Gaussian profiles to each `blob' in each
velocity channel in the PPV cubes of spectral lines.  Unlike previous works, we
have had to fit multiple Gaussians in several channels, since we resolve the
structure and see `blobs' both above and below the disk.  Figures
\ref{fig:velcen_U4}, \ref{fig:velcen_sio}, and \ref{fig:velcen_h2o} show the
results of this analysis.

We have modeled the velocity profile assuming an edge-on, uniform, optically
thin disk with a sharp central hole and outer truncation.  Position-velocity
curves derived with this approach are shown in the above figures. Figure
\ref{fig:massradiusdemo} shows the curves for a range of masses and radii.
This model approach is the same used by \citet{Plambeck2016a}.  We fit this
model to the centroid data points.  The fits were performed on the
\emph{average} positional offset at each velocity, since for many velocities
there were two or more Gaussian components fitted in the image.  The mass,
inner and outer radius, and centroid velocity were left as free parameters.
The fit results are shown in the legend of Figures 
\ref{fig:velcen_U4} and \ref{fig:velcen_sio}; in Figure \ref{fig:velcen_h2o},
we show only a fiducial model because the best-fit model did not describe
the data well.

The positions of the fitted components are significantly different for each
species, which helps illustrate why previous estimates of \sourcei's mass were
low.  Fits to both the SiO line in Figure \ref{fig:velcen_sio} and water in
Figure \ref{fig:velcen_h2o} have lower masses than the fit to the U232.511 line
in Figure \ref{fig:velcen_U4} that more closely traces the disk.

In the edge-on disk models, different inner-radius cutoffs have an effect on
the inner velocity profile slope similar to changing the central mass, so it is
likely that the disk parameters, rather than the central source mass, dominate
our uncertainties in this approach.  
Figure \ref{fig:massradiusdemo} demonstrates this effect: the inner slope of
a 5 \msun, $20~\mathrm{AU} < r < 50~\mathrm{AU}$ disk is indistinguishable
from a 20 \msun , $30~\mathrm{AU} < r < 80~\mathrm{AU}$ disk, though the latter
extends to higher velocity and radius.

\FigureTwo
{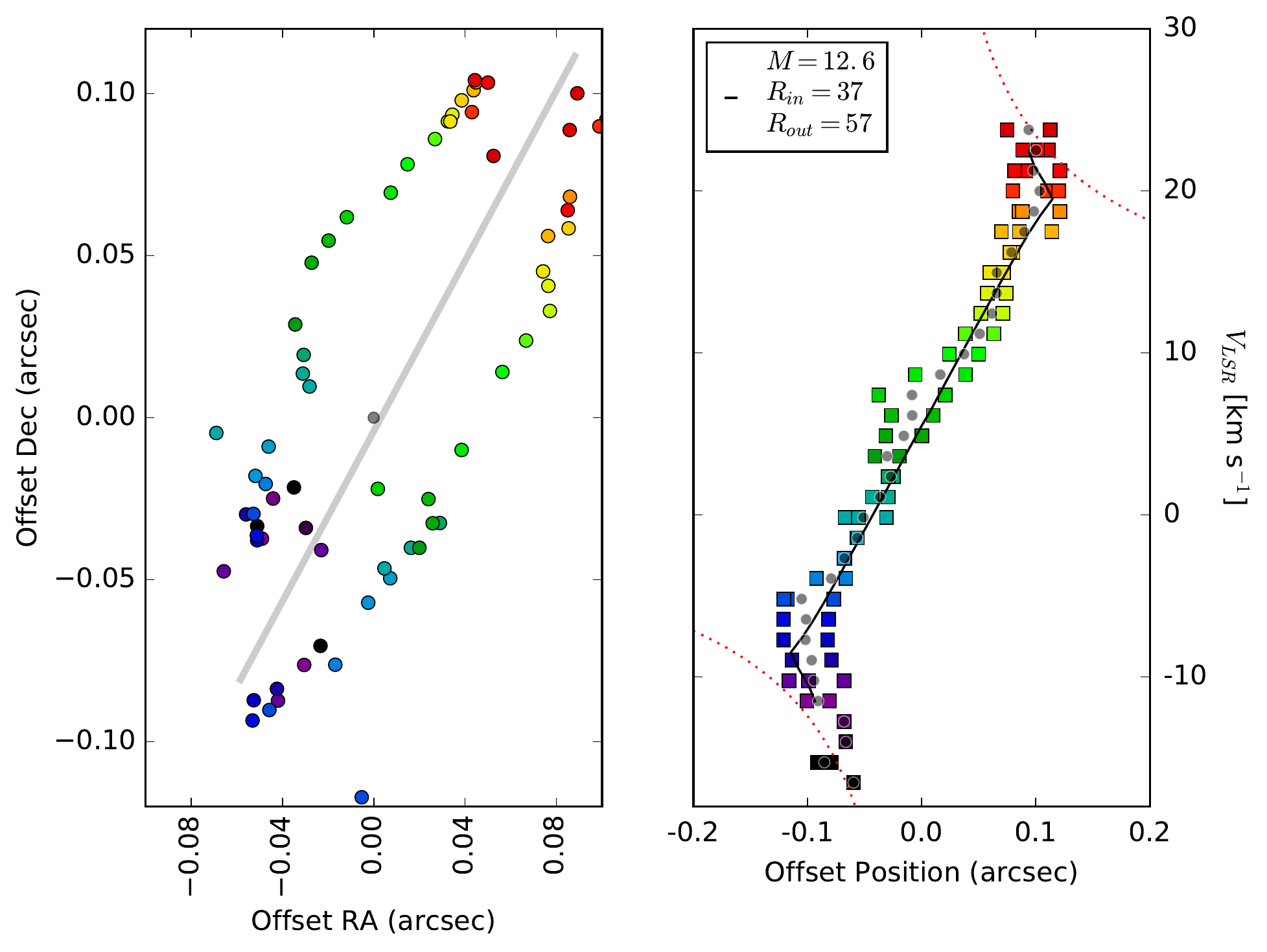}
{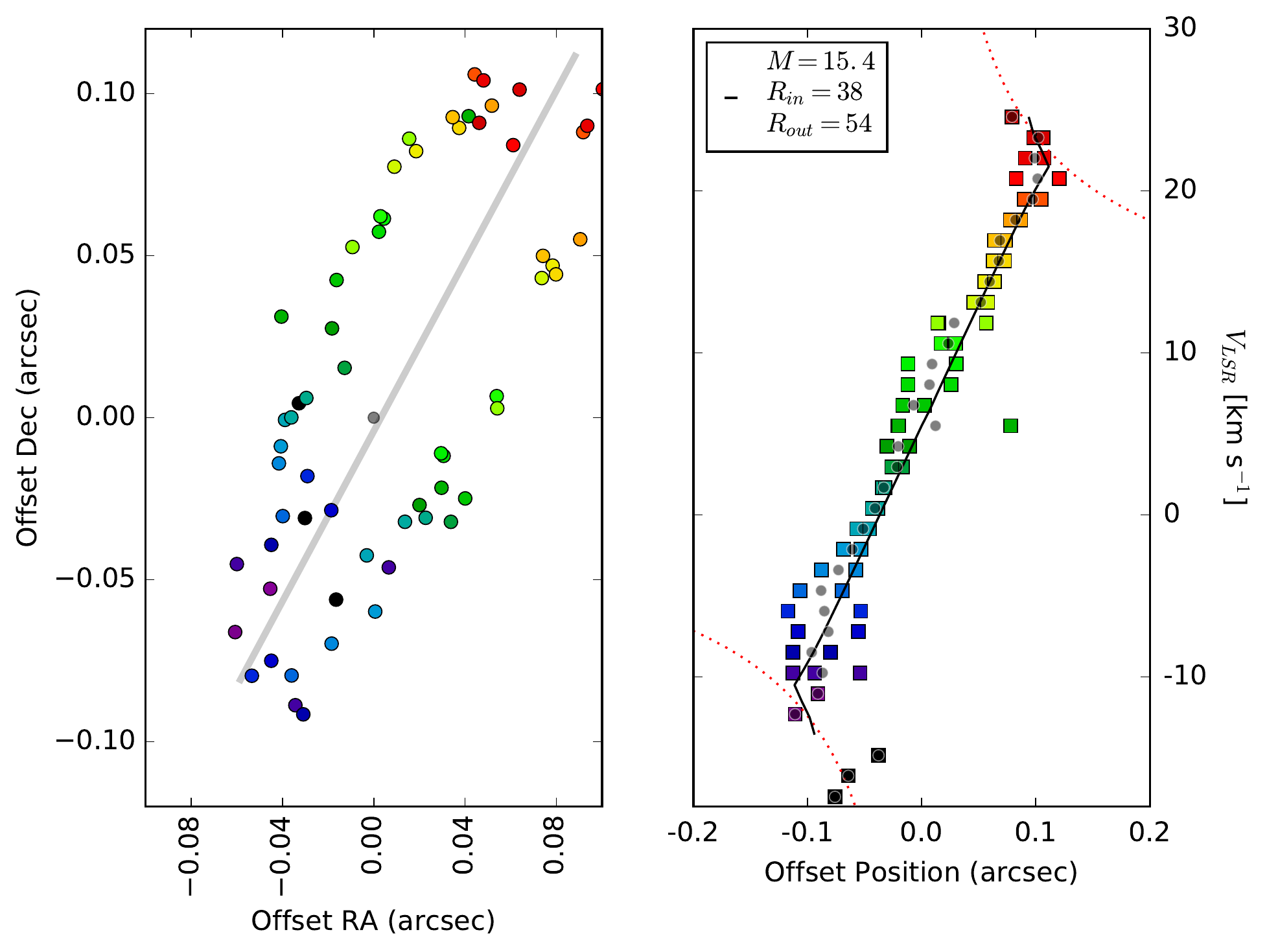}
{Results of the centroid-velocity analysis for the U232.511 line (left) and the U230.322 line (right).
The left panel shows the locations of fitted centroids in the position-position
plane relative to the midpoint of the disk.  
The position of the central compact source is
marked with a grey circle at the center.  The grey line indicates the disk
midplane as determined from the continuum modeling.  The circles are colored by
their velocity as indicated in
the right panel.  The right panel shows a position-velocity diagram of these
same centroids.  The dotted external curves show Keplerian velocity profiles
for a  15 \msun (red solid) central source; this curve does \emph{not} represent
what should be observed in a centroid-of-velocity plot.
The black curve shows the predicted centroid velocity profile of an
optically-thin edge-on disk with parameters displayed in the figure.
}
{fig:velcen_U4}{1}{3.5in}

\Figure
{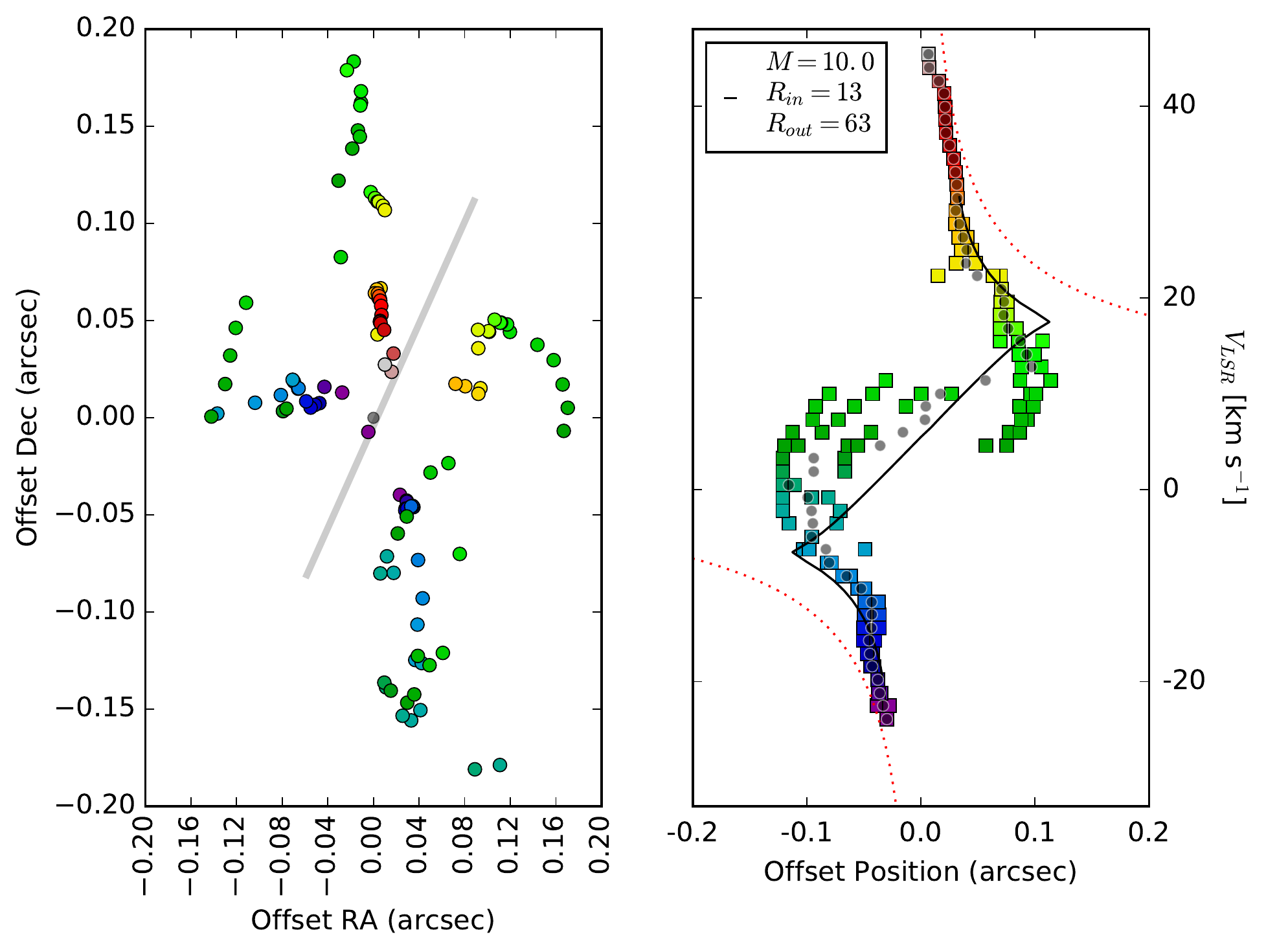}
{Results of the centroid-velocity analysis for the SiO v=1 J=5-4 line.
See Figure \ref{fig:velcen_U4} for details.
Note that the data are inconsistent with the disk model;
the SiO emission fitted here traces the bottom of the outflow
and possibly some components of the disk upper atmosphere.
The average positions at each velocity are closer to a reasonable fit.
}
{fig:velcen_sio}{1}{6in}

\Figure
{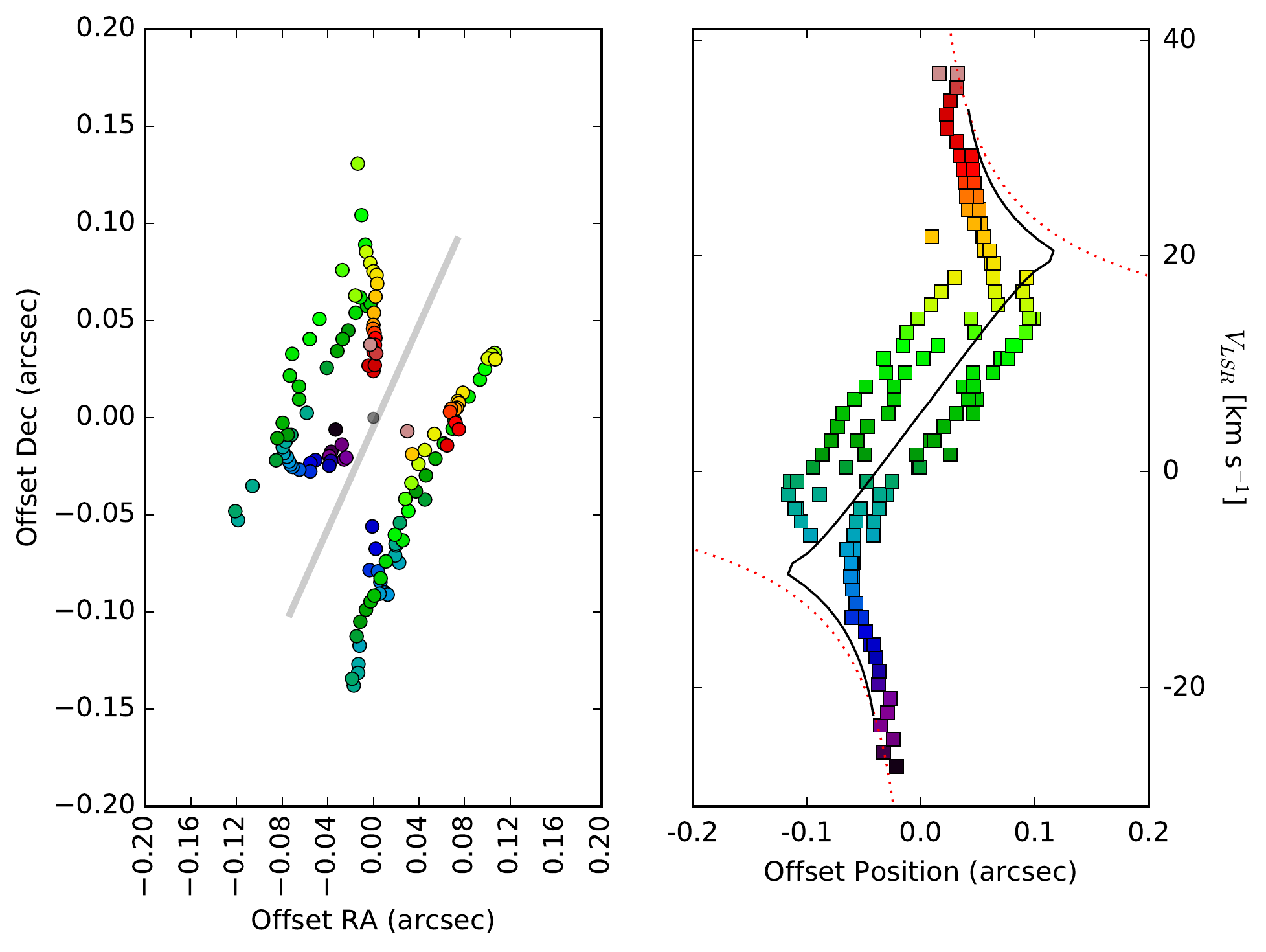}
{Results of the centroid-velocity analysis for the \water line.
See Figure \ref{fig:velcen_U4} for details.
The model disk fit to these data was a very poor fit, so we
have instead overlaid a model that is \emph{not} a fit to the data
with 15.5 \msun, $r_{inner}=17$ AU, and $r_{outer}=66$ AU in the
right panel.
Note that the centroid positions do not extend as far from
the source center as the unknown lines; this effect is 
a symptom of the blending of lines of sight in the centroid-based approach,
since the \water line's faintest emission can be seen extending to at least as
great a distance from the central source as the unknown lines in the
position-velocity diagrams.
}
{fig:velcen_h2o}{1}{6in}

\FigureOneCol{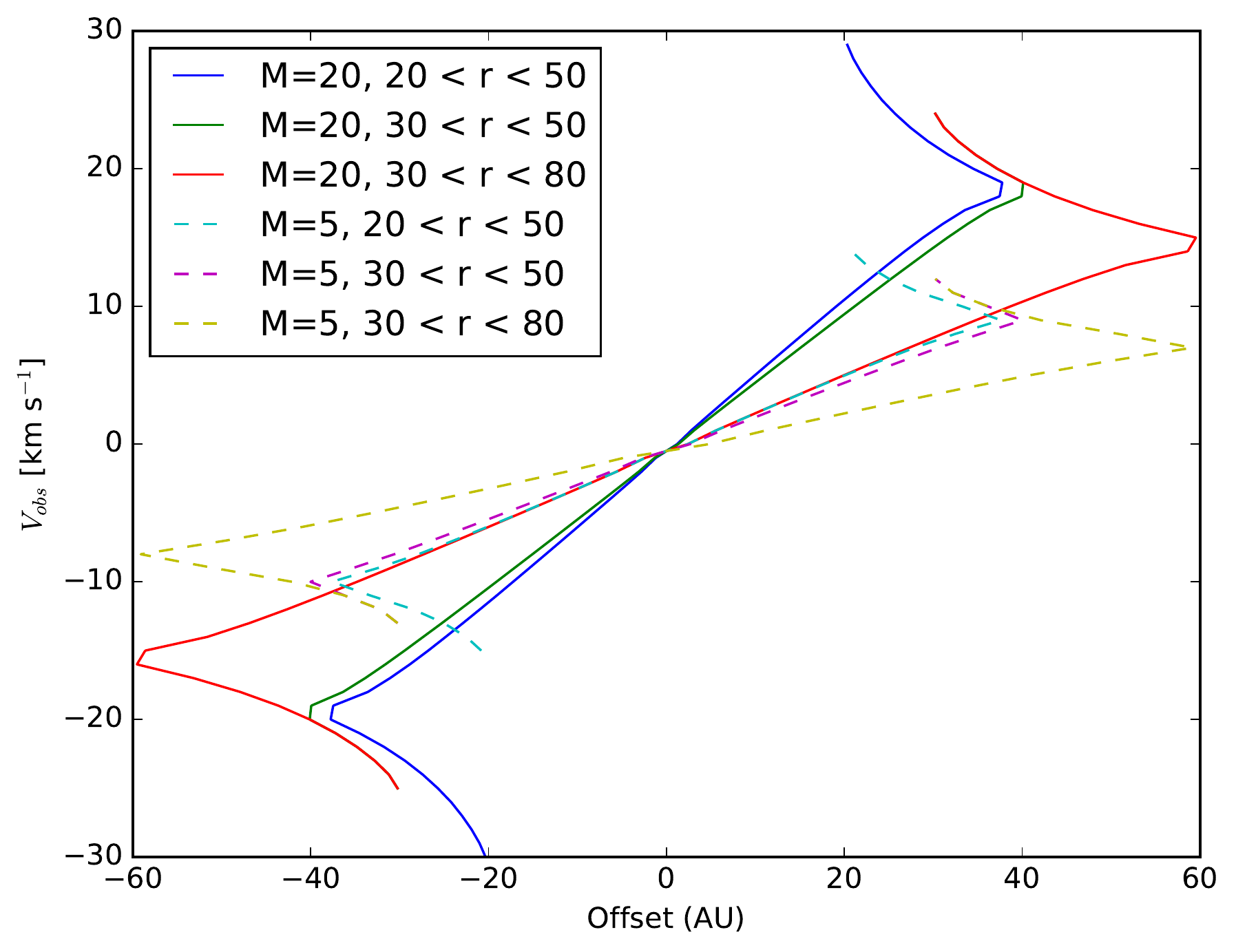}
{Plots of the predicted centroid rotation curves for different masses
and inner and outer radial cutoffs.  As discussed in Appendix \ref{appendix:centroids},
this figure illustrates the ambiguity between the radial extent of the disk and
the mass of the central source, since an $M=5$ \msun central source with a
$20<R<50$ AU disk has the same slope in the inner part as a $M=20$ \msun source
with a $30<R<80$ AU disk.
}
{fig:massradiusdemo}{1}{3.0in}

\subsection{A demonstration of issues with the centroid-of-velocity method}
Figure \ref{fig:h2opvfailuremode} shows an example of how the
centroid-of-velocity approach produces lower mass fits for some lines,
particularly \water and SiO.  The figure shows both the optically thin edge-on
disk model and the midplane-extracted position-velocity diagram with overlaid
centroid fits.  The centroid fits notably do not extend nearly as far as
emission is visible.  This discrepancy results from the midplane emission being
much fainter than some of the off-plane emission.

\Figure{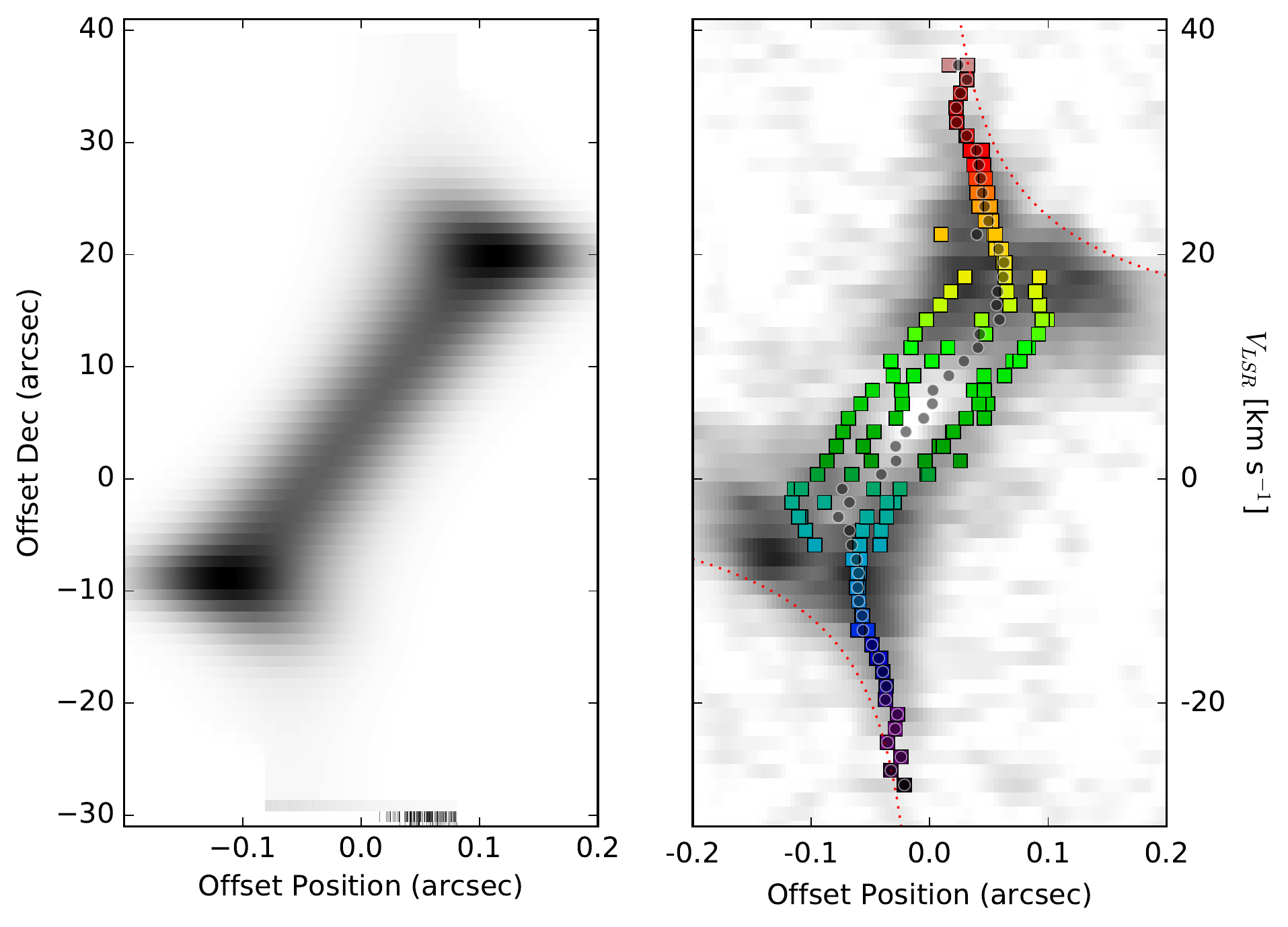}
{The edge-on disk model used to fit the centroid-of-velocity curves with
$M=15 \msun$, $r_{in}=25$ AU, and $r_{out}=65$ AU (left) and the centroid-of-velocity measurements
overlaid on the midplane position-velocity diagram of the \water line (right).}
{fig:h2opvfailuremode}{1}{6in}

By contrast, a similar side-by-side comparison of the edge-on optically thin
model with the U232.511 line reveals a better match.  In Figure
\ref{fig:u4pvmodelcomparison}, the overall structure of the observed
position-velocity diagram is well-matched to the model.

\Figure{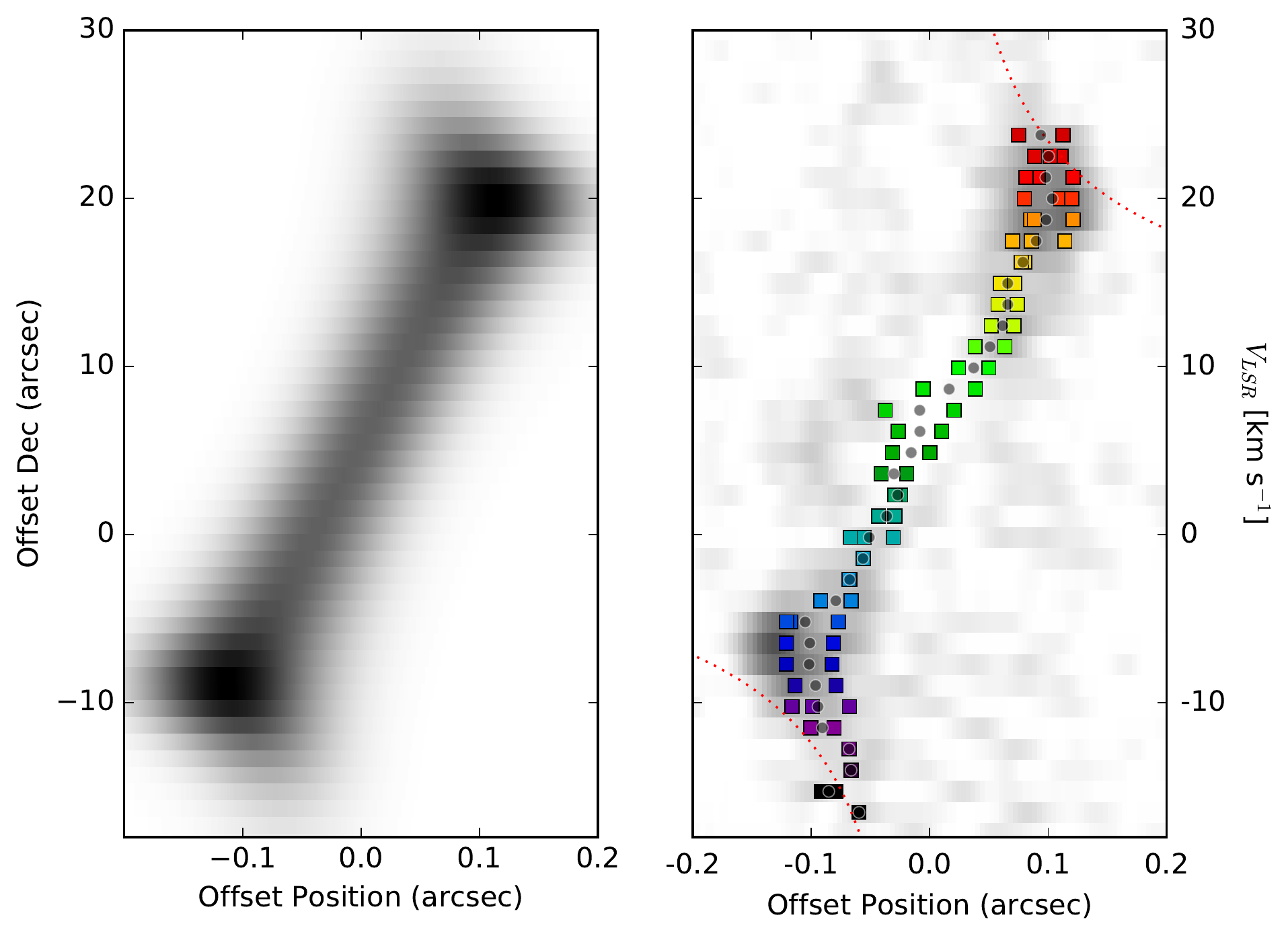}
{The edge-on disk model used to fit the centroid-of-velocity curves with
$M=15 \msun$, $r_{in}=25$ AU, and $r_{out}=65$ AU (left) and the centroid-of-velocity measurements
overlaid on the midplane position-velocity diagram of the U232.511 line (right).}
{fig:u4pvmodelcomparison}{1}{6in}

\section{Additional figures showing the disk}
\label{sec:ulinefigures}
We include several additional figures showing the disk moment 0 maps and
position-velocity profiles for some other unknown lines.  These figures show
that the lines displayed in the main text are not unique.

Figure \ref{fig:h2omom0} shows moment 0 maps, which provide a slightly
different view from the peak intensity maps shown in Figure \ref{fig:U1peak}.
Figures \ref{fig:U4} and \ref{fig:U5} show peak intensity, moment 0, and
position-velocity maps of the U232.511 and U217.980 lines.

\FigureThree
{{f39}.pdf}
{{f40}.pdf}
{{f41}.pdf}
{Moment-0 (integrated intensity) map of the U230.322 (left) and \water (middle,
right) line with continuum overlaid in contours.  Continuum contours
from the robust -2 map are shown
in red at levels of 
50,  300, and 500 K.
Contours of the line data are shown at 5, 10, and 15 $\sigma$.
The left two figures are robust 0.5 weighted images, while the right
is a robust -2 weighted image with higher resolution and poorer
sensitivity.  The U230.322 line is not detected in the robust -2 cubes.
}
{fig:h2omom0}{1}{2.2in}

\FigureThree
{{f42}.pdf}
{{f43}.pdf}
{{f44}.pdf}
{Moment 0 and peak intensity map of U232.511, similar to Figures \ref{fig:U1peak} and \ref{fig:h2omom0}.
The rightmost panel shows a position-velocity diagram extracted from the disk midplane.
The overlaid  curve shows the Keplerian velocity profiles for a 15 \msun central source in red.
The dashed magenta lines show the velocity curves at r=30 and 70 AU.
In the left and right panels, contours of the line data are shown at 5 and 10 $\sigma$,
and in the center panel, they are at 50, 100, 150, and 200 K.
}
{fig:U4}{1}{2.25in}

\FigureThree
{{f46}.pdf}
{{f47}.pdf}
{{f48}.pdf}
{Moment 0 and peak intensity map of U217.980, similar to Figures \ref{fig:U1peak} and \ref{fig:h2omom0}.
The rightmost panel shows a position-velocity diagram extracted from the disk midplane.
The overlaid  curve shows the Keplerian velocity profiles for a 15 \msun central source in red.
The dashed magenta lines show the velocity curves at r=30 and 70 AU.
In the left and right panels, contours of the line data are shown at 5 and 10 $\sigma$,
and in the center panel, they are at 50, 100, and 150 K.
}
{fig:U5}{1}{2.25in}

\end{document}